\documentclass[final,3p,times,twocolumn]{elsarticle}

\usepackage{amssymb}
\usepackage[utf8]{inputenc}
\usepackage[T1]{fontenc}

\usepackage{amsmath}
\usepackage{xcolor}
\usepackage{hyperref}
\usepackage{bm}
\usepackage{booktabs}     
\newcommand{\likeli}[2]{\mathcal{L}(#1 |\thinspace #2)}
\newcommand{\evid}{\mathcal{Z}}

\newcommand{\jm}[1]{\textcolor{purple}{#1$_{_{\mathbf{JM}}}$}}

\journal{New Astronomy Reviews}

\begin{document}

\begin{frontmatter}



\title{Weak in the Presence of Beauty: Gravitational waves from the mergers of black holes and neutron stars as a messenger}


\affiliation[label1]{organization={School of Physics, Mathematics and Computing},
            addressline={University of Western Australia}, 
            city={35 Stirling Highway},
            postcode={Crawley 6009 WA}, 
            country={Australia}}

\affiliation[label2]{organization={OzGrav: The ARC Centre of Excellence for Gravitational-wave Discovery},
            country={Australia}}

\author[label1, label2]{Fiona H. Panther} 
\author[label1, label2]{Jordan W. N. Moncrieff}
\author[label1, label2]{Mac L. Button}


\begin{abstract}
In 2025, the 4th Gravitational Wave Transient Catalog reported the discovery of over 200 gravitational wave events since 2015. These gravitational waves (GWs) are emitted by compact binary coalescenses (CBCs) between neutron stars and black holes, and detected with a global network of ground-based laser interferometers. GWs represent a new `messenger' -- ripples in spacetime that propagate across cosmological distances, and carry information about the physical properties of compact objects and their location in the Universe. In this review, we consider what information we can learn about black holes and neutron stars using GWs as a messenger. We explore how the GW observation process allows us to transform the outputs of laser interferometers like LIGO, Virgo and KAGRA into knowledge about black holes and neutron stars. We show that the GW source population acts as a messenger that can lead us to breakthroughs in our understanding of stellar evolution, how the Universe has changed across cosmic time, and fundamental physics. Finally, we discuss three key opportunities to learn more about our Universe using GWs as a messenger in the coming decades. 
\end{abstract}




\end{frontmatter}



\section{Introduction}
\label{sec:intro}

Although gravity is the weakest of the four fundamental forces, it shapes the Universe we observe today. In its Newtonian limit, gravity can explain the formation of large scale structure, binding together the largest structures in the Universe like galaxy clusters and the filamentary structure of the cosmic web. However, the General Theory of Relativity (`general relativity' or GR, \cite{EinsteinGR}) can explain far more about our Universe than Newtonian gravity, and is rich with predictive power when it comes to phenomena we can observe. The curvature of spacetime close to large masses like planets and stars can bend the path taken by rays of light from distant sources \citep[see ][for an overview]{Dyson1920}, and causes higher-order corrections to the orbits of planets \citep{Perihelion}. These initial predictions were rapidly validated in the early 20th century using light as a messenger. However, not all predictions of GR were so easily tested. Einstein observed that his field equations had wavelike solutions - small perturbations of spacetime that propagate at the speed of light that Einstein initially dismissed only as a mathematical curiosity. The work of Trautmann and Robinson, inspired by Infeld's effective field theory, identified that GWs must exist in nature \citep{Salisbury2019}. Confirmation of this work followed with observations of the Hulse-Taylor double pulsar \cite{hulse1975pulsar}. However, the evidence for GW emission was here only indirect, inferred from the shortening of the orbital period of the two bodies as conveyed through the time differences between successive pulses of electromagnetic radiation. Here, the prediction that gravitational waves must exist in our physical world was tested using light as a messenger. However, to truly test Einstein's theory, a direct observation of these ripples in spacetime was required.

Such a measurement came in 2015 when GW150914, the merger of two $\sim30\,\mathrm{M_\odot}$ black holes was detected \citep{gw150914} by the Laser Interferometer Gravitational wave Observatory (LIGO) \citep{LIGOScientific:2014pky}. Since then, the field of gravitational wave astrophysics has gone from the groundbreaking first discovery of a merging compact binary, to the detection of such events becoming commonplace \citep{GWTC4} with a global network of interferometers (LIGO, Virgo~\citep{VIRGO:2014yos, standard_virgo}, KAGRA~\citep{kagra_2020, kagra_2012, kagra} and until recently, GEO600~\citep{standard_geo}). 

A single gravitational wave event acts as a messenger, carrying with it information about the two merging compact objects (COs) (or any other physical system with a time-varying quadrupole moment). From an observation of a gravitational wave, we can infer the properties of black holes and neutron stars from billions of light years away. A small fraction of these events are expected to emit detectable electromagnetic radiation, when matter is accelerated close to the speed of light in the extremes of curved spacetime created by merging neutron stars. This was the case with the detection of the first binary neutron star merger GW170817 \citep{GW170817} - an event that allowed us to view the moments before and during the merger through the emission of gravitational waves. The electromagnetic emission associated with GW170817 allowed us to study the behavior of the matter left behind. This formed a radioactively-powered kilonova \citep{metzger10} and a relativistic jet. Such multimessenger events offer prized opportunities beyond merely measuring the properties of black holes and neutron stars, acting as beacons to enable precision measurements of the expansion rate of the Universe by exploiting the `standard siren' property of GWs \citep{hughes05}. 

With over 200 compact binary coalescences (CBCs, the collective term for binary neutron star, neutron star-black hole, and binary black hole mergers) detected and reported in the 4th Gravitational Wave Transient Catalog \citep{GWTC4}, we now have more power than ever to use GWs as a messenger to glean information about the Universe. The GW population acts as a messenger, one that can only emerge once hundreds of events have been detected \citep{gwtc4pop}. This is the era we now enter in GW astrophysics: one where population-level inferences can be made about stellar evolution, compact object formation, cosmology and fundamental physics. These inferences make a GW source population more than merely the sum of its parts. 

This review will concentrate on gravitational wave transients observed by ground-based gravitational wave detectors: CBCs where emission can be characterized by some finite start and/or end time (whether that is in passing through the LIGO frequency band, or the emission is halted by some physical process). It is set out as follows: We will first review how gravitational waves can convey information about their sources in Section \ref{sec:waves}. In Section \ref{sec:detection} we will review the methods of gravitational wave detection, with a particular focus on the relationship between GW as a carrier of astrophysical information, and how the detection process allows us to optimally extract this information. In Section \ref{sec:pops}, we will review how the population of CBC sources acts as a messenger about the astrophysics of compact binary object formation. In Section \ref{sec:nonvanilla}, we see how individual GW events exhibit higher-order effects that allow us to more definitively probe hypotheses about compact binary formation and the General Theory of Relativity. In Section \ref{sec:multi} we discuss searches for coincident electromagnetic emission from merging compact binaries, and the successes and challenges presented by this work. We investigate how GWs can serve as a messenger of fundamental physics in Section \ref{sec:fundamental}. Finally, we discuss two potential avenues for GWs as a messenger of astrophysics in Section \ref{sec:future}, motivated by the work of the authors of this review.  

\section{Gravitational Waves as a messenger}\label{sec:waves}
\begin{figure}[h]
\centering
\includegraphics[width=\columnwidth]{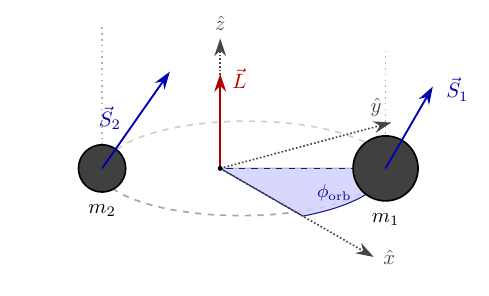}
\caption{Basic properties and geometry of a compact binary system}\label{fig:system}
\end{figure}

\begin{figure}[h]
\centering
\includegraphics[width=\columnwidth]{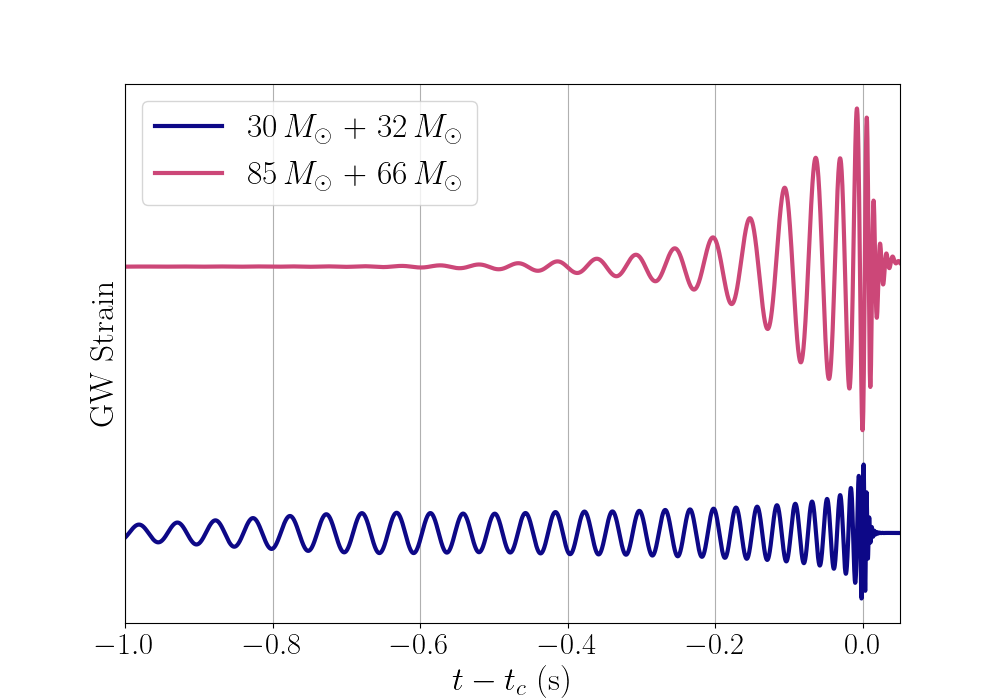}
\caption{Gravitational waveforms from non-spinning binary black holes with different chirp masses. Larger chirp mass systems produce shorter signals with lower merger frequencies. The chosen component masses represent the gravitational waves detected from events GW150914 and GW190521, simulated using \texttt{IMRPhenomXPHM}\citep{Pratten2021} using the \texttt{PyCBC} software library \citep{PyCBC}}\label{fig:bbhcomp}
\end{figure}

For a binary system (Fig \ref{fig:system}) composed of two compact objects with masses $m_1 + m_2 = M$ located at a distance $d_L$ and chirp mass \citep{Peters1963}
\begin{equation}
    \mathcal{M} = \frac{(m_1 m_2)^{3/5}}{(m_1 + m_2)^{1/5}},
\end{equation}
the gravitational wave strain can be expressed as the sum of two independent and orthogonal polarisations $h(t) = h_+(t) + h_\times(t)$ in the time domain as 
\begin{align}\label{eq:wavefull}
    h_+ &=A(t) \bigg(\frac{1+\cos^2\iota}{2}\bigg)\cos 2\phi_\mathrm{orb}(t),\\
    h_\times &=A(t)\cos\iota \sin2\phi_\mathrm{orb}(t).
\end{align}
Where $A(t)$ is given by
\begin{align}
    A(t) &= \frac{G\mathcal{M}}{d_Lc^2}\bigg(\frac{t_c- t}{G\mathcal{M}/c^3}\bigg)^{-1/4}.
\end{align}
$t_c$ is the time of coalescence, $G = 6.67\times10^{-11}\,\rm{m^3 kg^{-1}s^{-2}}$ is the gravitational constant and $c$ is the speed of light. The inclination angle $\iota$ is the angle of inclination of the orbital plane relative to the observer. We can easily see from Equation \ref{eq:wavefull} that the frequency evolution and duration of the waveform observed by ground-based GW detectors is related to the chirp mass of the system: more massive binary systems produce larger amplitude and lower frequency signals than low mass binaries at the same luminosity distance, and merge at lower frequencies (Fig. \ref{fig:bbhcomp}). 

The term $\phi_\mathrm{orb}$ is the angle between $m_1$ and the x-axis at some time $t$ (Fig. \ref{fig:system}). Energy loss from the system through the emission of GWs leads to the time evolution of $\phi_\mathrm{orb}$. It is possible to express the orbital phase in terms of the gravitational phase $\phi$ as 
\begin{equation}
    2\phi_\mathrm{orb} = \phi(t) + \phi_c
\end{equation}
where $\phi_c$ is the phase at the time of coalescence $t_c$. Hence, $\phi(t)$ contains all the information about the time evolution of the phase due to GW emission and general-relativistic corrections to the orbits. For example, at second post-Newtonian order, the orbital phase of a non-spinning binary evolves as
\begin{multline}\label{eq:orb}
    \phi(t) = -\frac{2}{\eta}\bigg\{\Theta(t)^{5/8}+ \bigg(\frac{3715}{8064}+\frac{55}{96}\eta\bigg)\Theta(t)^{3/8} \\
    - \frac{3\pi}{4}\Theta(t)^{1/4}+ \bigg(\frac{9275495}{14450688} + \frac{284875}{258048}\eta+\frac{1855}{2048}\eta^2\bigg)\Theta(t)^{1/8}\bigg\}
\end{multline}
where $\eta = \mu/M$ is the symmetric mass ratio and  $\phi_c$ is the orbital phase at $t=t_c$ and 
\begin{equation}
    \Theta(t) = \frac{c^3\eta(t_c - t)}{6GM}
\end{equation}
is the dimensionless time parameter. A measurement of the gravitational wave strain $h$ therefore conveys information about the source of the gravitational waves: both intrinsic properties such as the masses and mass ratio of the merging compact objects, and the extrinsic properties of the binary system such as its inclination angle and distance from the observer. From Equation \ref{eq:orb} we observe that higher order corrections to the waveform will be most readily detectable through the time evolution of the orbital phase for highly asymmetric systems.

The post-Newtonian (PN) relativistic corrections that affect the phase evolution of the merging binary can be explored through numerical relativity simulations. Restricted post-Newtonian waveforms that incorporate corrections in the orbital phase such as in Equation \ref{eq:orb} have proven to be more than adequate for searches for compact binary coalescenses and, in many cases, parameter estimation. Both of these tasks require parametric waveform models that can be evaluated with minimal computation time. A comprehensive discussion of computing post-Netwonian corrections and applications of numerical relativity simulations to compute high-precision waveform models can be found in \cite{Buonanno2007}. 

\subsection{The effects of spin, eccentricity, and tidal deformability}
Astrophysical BHs are characterized by their mass and their spin $\bm{S}$ (Fig \ref{fig:system}). For each compact object in a merging system, the spin is often expressed in terms of the dimensionless spin, 
\begin{equation}
    \bm{\chi}_i \equiv \frac{c\bm{S}_i}{Gm_i^2}.
\end{equation}
The orientation and magnitude of the BH spins are determined by processes that occur during stellar evolution, and dynamical interactions between the black hole and its environment. We explore the importance of spin in constraining the processes that occur during BBH formation and evolution in Section \ref{sec:pops}. The inspiral rate (and hence phase evolution) of the inspiral is determined by the mass ratio, as above, with an additional contribution from the spins parallel to the orbital angular momentum (e.g. see \citep{Buonanno2007}). The effect on the waveform can be quantified using the effective spin parameter $\chi_\mathrm{eff}$ \citep{Darmour2001}
\begin{equation}\label{eq:chieff}
    \chi_\mathrm{eff}= \frac{(\bm{\chi_1} + q\bm{\chi_2})\cdot \hat{\bm{L}}}{1+q},
\end{equation}
where $\hat{\bm{L}}$ is the unit vector in the direction of the orbital angular momentum $\bm{L}$ and $q = m_2/m_1$ is the mass ratio of the two COs. Positive values of $\chi_\mathrm{eff}$ indicate that black hole spins are net aligned with the orbital angular momentum (`prograde'), and negative values indicate a net anti-alignment (`retrograde'). Small values of $\chi_\mathrm{eff}$ are consistent with either small and randomly-oriented spins, or spins that lie within the orbital plane (both of which render the dot product in Eq. \ref{eq:chieff} small). As $\chi_\mathrm{eff}$ appears at low orders in post-Newtonian corrections to the waveform phase, and because it is a constant of the motion at second PN order, it can be well constrained in comparison to the individual component spins \citep{Gerosa2015}. In population analysis, measurements of the distribution of $\chi_\mathrm{eff}$ are used to constrain CO formation scenarios \cite{Colloms2025}, as isolated binary formation and dynamical formation result in different distributions of $\chi_\mathrm{eff}$ due to the assumption that BH spins tend to align during isolated stellar evolution \citep[although see ][and references therein for a discussion as to why this assumption may be flawed.]{Baibhav2024}. 
\begin{figure}[h]
\centering
\includegraphics[width=\columnwidth]{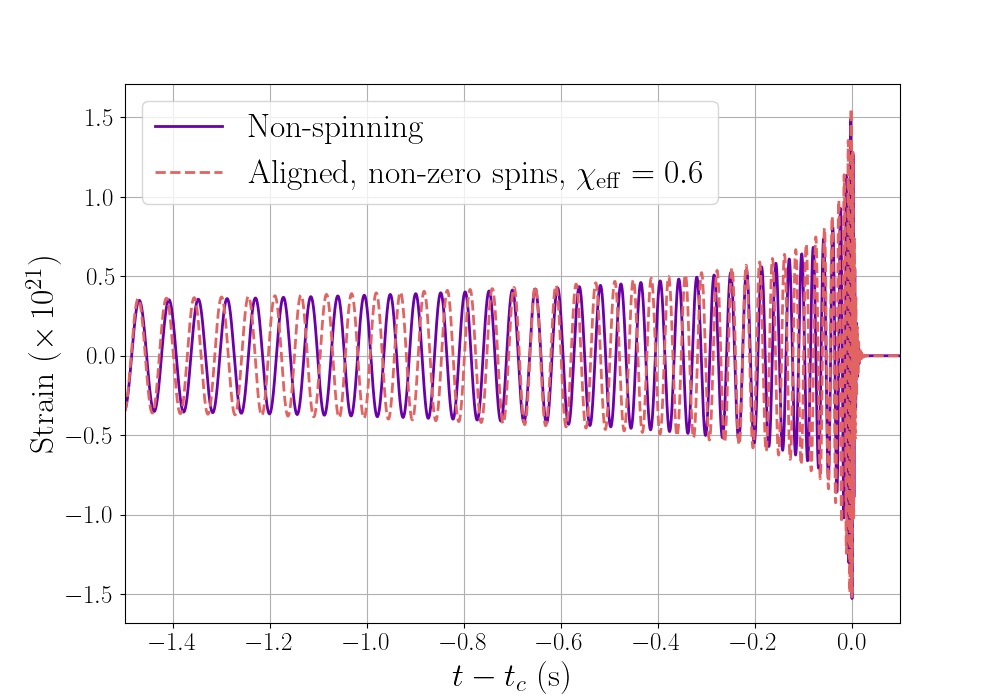}
\caption{BH spin induces additional phasing of $\phi_\mathrm{orb}$ during the inspiral. A 30+32$M_\odot$ non-spinning system (solid line) is shown in comparison to a system with non-zero spins aligned with $\vec{L}$, where $\chi_{1z} = 0.9,\,\chi_{2z}=0.3,\,\chi_\mathrm{eff}\approx 0.6$. Waveforms generated using the \texttt{IMRPhenomXPHM} waveform approximant \citep{Pratten2021} and the \texttt{PyCBC} software library \citep{PyCBC}.}\label{fig:bbhspin}
\end{figure}

In the case considered in Figure \ref{fig:bbhspin}, the black hole spins are aligned with the orbital angular momentum of the system. However, if the spins of the individual black holes are misaligned with the orbital angular momentum (as might be expected if black hole binaries form through dynamical capture), the misalignment results in the precession of both the orbital plane and the spin vectors. 

The magnitude of the spin precession induced by these misalignments leads to additional amplitude and phase modulations in the waveform with respect to the spin-aligned case. By averaging the time-dependent effects of spin precession \citep[e.g. see][]{Schmidt2015}, we can define the effective spin precession parameter
\begin{equation}
    \chi_p = \mathrm{Max}\bigg(\chi_1\sin\theta_1, \bigg(\frac{4q+3}{4+3q}\bigg)q\chi_2\sin\theta_2\bigg), 
\end{equation}
to quantify the degree to which precession impacts the phase and amplitude modulation of the waveform. Here, $\theta_i$ is the angle between the orbital angular momentum vector $\vec{L}$, and the spin vector $\vec{S}$ of BH $i$. Two phenomena contribute to the modulation of the waveform and to the measurable value of $\chi_p$: the magnitude of the spins in the orbital plane oscillate around a mean value, and the relative angle between the spin vectors in the plane change continuously as the black holes orbit. The effect of spin precession is more pronounced for unequal mass binaries where $q\ll1$ - searches for evidence of spin precession have therefore focused on the most unequal-mass systems identified by the LVK (see Section \ref{sec:spinprec}). However, the effects of spin precession are easily confounded by the presence of noise transients, and it is challenging to distinguish between the effects of spin precession and binary eccentricity.  

\begin{figure}[h]
\centering
\includegraphics[width=\columnwidth]{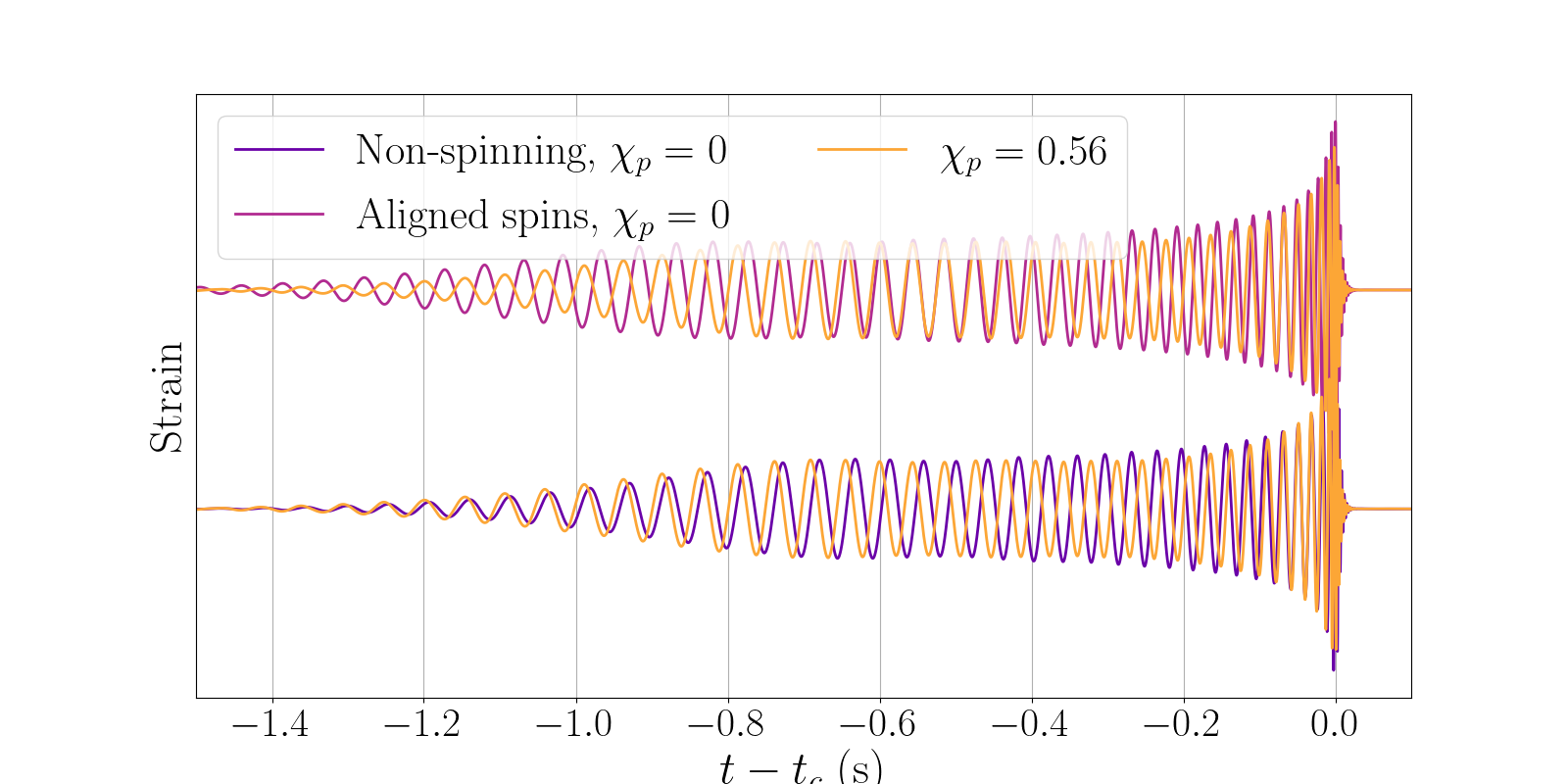}
\caption{Comparison between waveform produced by a spin-precessing BBH binary with the same masses in Fig. \ref{fig:bbhspin} ($\chi_p=0.56$) with a non-spinning binary (bottom trace, solid) and a spin-aligned, non-precessing binary (top trace). The precession of the orbital plane causes additional amplitude and phase modulations relative to both the non-spinning and spin-aligned waveform. Waveforms generated using \texttt{IMRPhenomXPHM} and \texttt{PyCBC} \citep{Pratten2021, PyCBC}}\label{fig:bbhsignal}
\end{figure}

\subsection{Binary eccentricity}
A further phenomenon that has a modulating effect on the inspiral waveform's amplitude and frequency is that of binary eccentricity. Substantially eccentric binaries display substantial amplitude modulation due to the close approach between the bodies at pericenter releasing orbital energy from the system at an enhanced rate compared to a circularised binary. Relativistic effects also lead to apsidal precession of the orbit. These effects compound, causing the orbit to circularise over many periods. This means that eccentricity that is detected during the last few moments of the inspiral provides compelling evidence for relatively recent dynamical interactions, as the system has not had time to circularise. A measurement of non-zero eccentricity is therefore a smoking gun for the dynamical formation of black holes.  

Observationally, eccentric mergers exhibit qualitatively similar amplitude and phase modulations as precessing systems and can be hard to distinguish \cite{bustillo2021confusing}. There have been numerous efforts to develop eccentric waveform models (summarised in references in \cite{Shaikh2023}) that can be efficiently computed to incorporate them into parameter estimation efforts \citep[Models such as \texttt{SEOBNRv4EHM} and \texttt{TEOBResumS-Dalí}; ][]{RamosBuades2022, Albanesi2025}. One complicating factor that makes eccentricity challenging to understand is the lack of a standardised definition of eccentricity in the General Theory of relativity that can be propagated into waveform models \citep{Shaikh2023, Shaikh2025}, and describing eccentric orbits requires the introduction of three additional parameters to describe the orbits of the COs: eccentricity, mean anomaly, and either the true anomaly or an alternative parameterisation to describe the size of the orbit. 

Eccentricity is difficult to measure at current detector sensitivities, but will be a key observable for next-generation ground- \citep{CEReitze} and space-based \citep{LISA_LRR,li2021Tianqin} gravitational wave detectors. Future ground-based detectors will probe earlier inspiral stages where residual eccentricity is more readily measured, while space-based detectors such as LISA are sensitive at lower frequencies, enabling observations of binaries long before merger when eccentricity is largest \cite{moncrieff2026roads}. Highly eccentric systems may also produce periodic bursts of higher-frequency emission during pericenter passage \citep{kocsis2012}, offering a potential multiband synergy between ground- and space-based observatories.

LISA will additionally be sensitive to extreme mass ratio inspirals (EMRIs) — stellar-mass black holes merging with supermassive black holes — at frequencies well below ground-based detector bandwidths, where eccentricity serves as a key discriminant between formation channels such as loss-cone scattering and disk-driven migration \citep{pan2021emri, mancieri2026eccentric}. More broadly, the distinct eccentricity distributions predicted by different formation scenarios — including BBH–SMBH triples \citep{grishin26eccentric} and various dynamical channels — make eccentricity a powerful probe of binary formation across detector bands \citep{LISA_LRR}.

\subsection{Tidal deformability}
The effects of spin on the inspiral waveforms emitted by merging neutron stars are less severe - binary neutron stars (BNSs) are thought to generally have small spin magnitudes as only around 4\% of systems are expected to have $\sim \,\mathrm{ms}$ spin periods \citep{Rosswog2023}. Although BNS spins are thought to be small, they can affect the overall inference of masses \cite{FarrBNS}, as high and low spin priors yield subtle differences in the measured component masses of events like GW190425 \citep{GW190425}. Merging BNS systems are also expected to almost exclusively arise from isolated binary evolution - simulations of dynamical compact object formation suggest that less than a few percent of low-mass CO binaries form through gravitational capture \citep{sam18, tagawa2020formation}. The emission of GWs during the early inspiral ($f_\mathrm{GW}\ll 10\,\mathrm{Hz}$ is also expected to damp any eccentricity the BNS system retains from processes that occur during isolated stellar evolution \citep{Dominik_2012,Mandel2016, Tagawa2018} For this reason, BNS systems are expected to have aligned spins and circularized orbits. There are, however, other important physical properties of neutron stars that can affect the phase and amplitude evolution of the late BNS inspiral (i.e. observed at frequencies $f>10\,\mathrm{Hz}$) that means waveform models must encompass additional parameters. 

Unlike black holes, neutron stars can be tidally deformed. The extent to which a neutron star can be tidally deformed depends strongly on the neutron star matter equation of state (EoS) which fixes the neutron star radius $R$ for a given mass $m$ \citep[see][for a review]{Lattimer2001}. In the context of GW emission, the phase and amplitude evolution of the inspiral signal depends on the tidal Love number $\lambda_i = Q/\epsilon$ where $Q$ is the tidally-induced quadrupole deformation of the $i$th star and $\epsilon$ is the tidal potential induced by the binary companion \citep{Binnington2009}. The dimensionless tidal deformability of the $i$th neutron star is given by \citep[e.g. ][]{Raithel2018}
\begin{equation}
    \Lambda_i = \frac{\lambda_i}{m_i^2} =  \frac{2k_{2, i}}{3}\bigg(\frac{c^2R}{Gm_i}\bigg)^{5}.
\end{equation}
Because the term $k_2, i$ is an equation-of-state and compactness dependent parameter \citep{Flanagan2008}, allowing a measurement of $\Lambda_i$ from the waveform to be transformed into an insight into the EoS \citep{Katerina2020}. The tidal deformation of the neutron stars accelerates the merger, and its effects on the frequency evolution of the waveform can be used to quantify the effective tidal deformability ($\tilde{\Lambda}$) of the binary system \citep{Hinderer2010}. The effective tidal deformability is a dimensionless combination of the component masses and tidal deformabilities, and is given by \citep[e.g. ][]{Raithel2018} 
\begin{equation}
\tilde{\Lambda} = \frac{16}{13}\bigg(\frac{(m_1 + 12m_2)m_1^4\Lambda_1 + (m_2 + 12 m_1)m_2^4\Lambda_2}{(m_1 + m_2)^5}\bigg).
\end{equation}
Depending on the accuracy with which the BNS component masses and spins can be measured, a measure of the NS tidal deformability with GWs can be converted into a strong constraint on the neutron star EOS and measurements of the neutron star radius to within $\delta r_\mathrm{NS}\sim 1\,\mathrm{km}$ \citep{Read2009}.

Tidal deformability is also measurable from NSBH waveforms in-principle. However, it is expected that most NSBH mergers do not involve the tidal disruption of the NS component unless the mass ratio of the system is small ($q<1/6$), the black hole spin is significantly prograde ($\chi_1\geq0.5$) or the neutron star radius is large (potentially implying a soft NS EOS) \citep{Kyutoku2011, Foucart2012, Hannam2013}. Systems comprising these properties are likely challenging to achieve in nature unless binary systems undergo mass ratio reversal, or the star that leaves behind the NS component evolves more rapidly, leaving the progenitor of the BH companion to be spun up. Nevertheless, identifying one component of a putative NSBH merger to have a non-zero tidal deformability (or, more specifically, $k_2>0$) provides the most concrete evidence that there is a neutron star present in the binary. Waveform models that incorporate this effect are based on NR simulations, or can be emulated with phenomenological modelling that truncates the waveform at frequencies of $\sim1500\,\mathrm{Hz}$, corresponding to the last stable orbit before the NS is tidally disrupted \citep[e.g. ][]{Clarke2023}.

\subsection{Higher order multipoles}
\begin{figure}[h]
\centering
\includegraphics[width=\columnwidth]{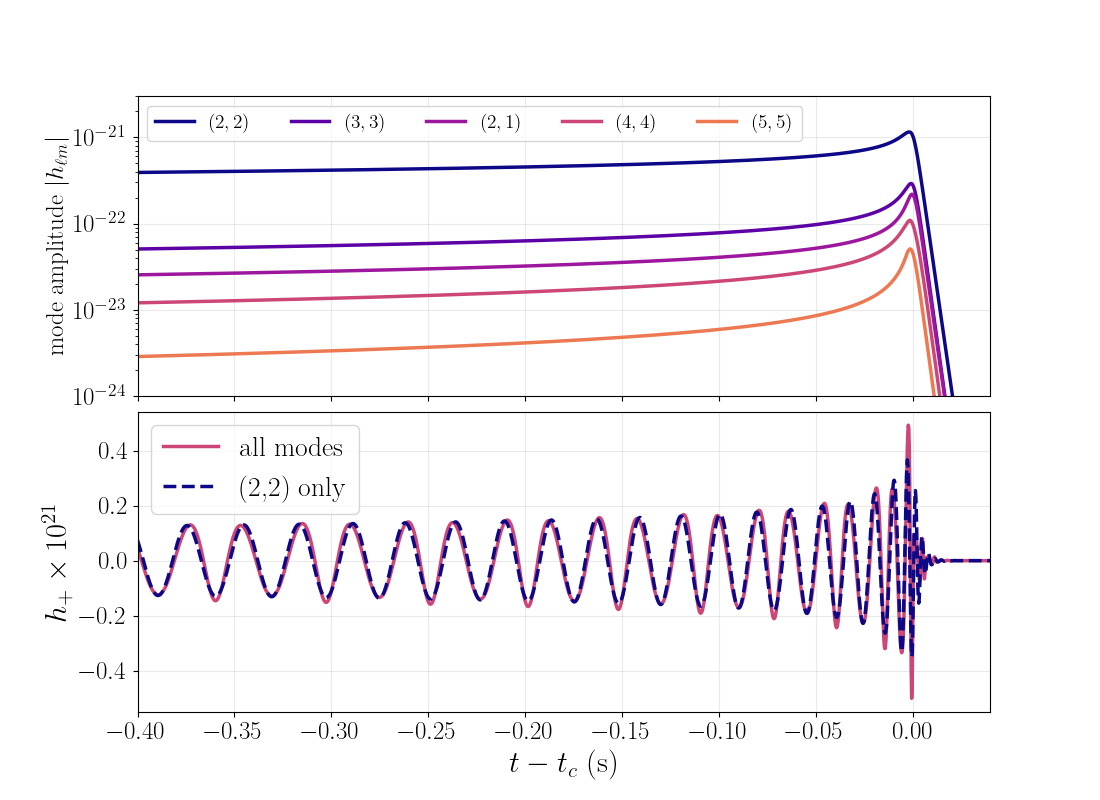}
\caption{Higher order multipoles generated by a merging BBH system with $m_1 = 40\,M_\odot, m_2 = 10\,M_\odot$ and $\iota\sim80\deg$. Higher order multipoles from high-mass ratio, edge-on CBCs make a non-negligible contribution to the overall detected signal-to-noise ratio. The strain amplitude of each of the multipoles is shown in the top panel. The overall observed waveform is shown in the lower panel. The visible mis-match between the sum of all multipolar modes and the (2,2) mode shows the importance of including higher-order multipoles in the study of CBCs with large mass ratios. Waveforms are geenrated with the \texttt{SEOBNRv4HM} waveform model \citep{Cotesta2021} and \texttt{PyCBC} \citep{PyCBC}}\label{fig:modes}
\end{figure}

 GW emission is quadurpolar. Consequently, the fundamental (2,2) quadrupolar mode dominates the observed GW signal at current detector sensitivities. Higher order multipole moments also contribute to the energy loss from the system (Fig. \ref{fig:modes}), especially if a system is oriented edge-on with respect to the observer. Consequently, the signal-to-noise ratio (and hence observability) of these higher harmonics in the inspiral depends on a number of factors, including the orientation of the binary, the total binary mass, and the binary mass ratio \cite{Mills2021}, . The observation of these higher order harmonics adds to the available evidence that the General Theory of Relativity is an accurate description of gravitation in extreme spacetime conditions. Through advances in numerical relativity and semi-analytic modelling, there is now an abundance of models that describe the contributions of higher harmonics to the GW signals we can observe with ground-based detectors \citep{Cotesta2021}. This has enabled the detection of signals that exhibit strong evidence for the presence of higher order harmonics (see Section 5.3), although these observations can be confounded by non-stationary noise in detectors in much the same way detection of precession and eccentricity is challenging. Non-stationary noise can induce variations in the phasing and amplitude modulation of the overall detected signal that are indistinguishable from higher order multipoles at current detector sensitivity. Beyond validating GR, observation of higher order modes can help in breaking the $\iota-d_L$ degeneracy that is present in the $(2,2)$ quadrupole mode. This is because the directional dependence of the emission of the $(2,1), (3,2), (3,3)$ and $(4,4)$ harmonics relative to the orbital plane allows the discrimination of the inclination angle of the binary $\iota$ \citep{Usman_2019}. This has important consequences for gravitational wave cosmology, as it can be used to reduce uncertainties in $d_L$ that propagate into the inference of the expansion rate of the Universe.

\section{The detection of gravitational waves with ground-based laser interferometers}\label{sec:detection}
Modern gravitational wave detectors are based on the Michelson interferometer. The mirrors of a gravitational wave interferometer can be suspended in such a way that they occupy an inertial reference frame, making them `test masses' that are freely falling, independent of the local gravitational potential of the Earth. Any change in the relative distance between the central beam splitter and the test masses (for example, the tiny displacements induced by a passing gravitational wave) causes changes in the interference pattern measured at the output photodetector. The measurable length change $\Delta L $ induced by a gravitational wave depends on the total length of the interferometer arms. In the limit that the arm length $L\ll\lambda_\mathrm{GW}$, the gravitational wave strain can be measured to first order as
\begin{align}
h(t)\approx \frac{\Delta L (t)}{L}   
\end{align}
where $h(t)$ is the dimensionless gravitational wave `strain'.

There are numerous sources of frequency-dependent (colored) noise in gravitational wave detectors that affect the precision with which we measure the arm length $L$ over time \citep{Freise}. This includes Newtonian noise at low frequencies from seismic disturbances, thermal noise from vibration of the mirrors and suspension system that alter arm lengths $L$, and quantum noise that arises from the photodetector. Each of these noise sources follow a stationary Gaussian process and can hence be characterised by the one-sided noise power spectral density $S_n(f)$, where
\begin{equation}
    2\mathbb{E}( \tilde{n}(f)^*\tilde{n}(f')) \equiv S_n(|f|)\delta(f-f') 
\end{equation}
where $\delta(f)$ is the Dirac delta function. It is worthwhile to be sure of the convention being used for the calculation of the PSD in gravitational wave applications, as there may be inconsistencies in whether the one- or two-sided PSD is used.

A GW interferometer is not uniformly sensitive to gravitational waves from all sky locations - the orientation of the detector arms relative to the gravitational wave source modulates the sensitivity of the detector according to the antenna response function. The observed strain $h^\mathrm{obs}(t)$ measured by the interferometer can be written as 
\begin{align}
    h^\mathrm{obs}(t) = F^+h_+(t) + F^\times h_\times(t)
\end{align}
where $F^{+/\times}$ depend on the three angles defining the sky position and polarization basis of the gravitational wave source with respect to a reference coordinate system. For example, when using equatorial coordinates they are a function of right ascension, declination and polarization angle as well as the detector location and the sidereal time. Most computer packages designed for detection and parameter estimation of gravitational wave signals \citep[e.g. ][]{lal, PyCBC, ashton19, gwbench} offer routines to directly compute the antenna pattern functions. GW detectors are most sensitive to sources that are directly above or below the plane that is defined by the plane spanned by the detector arms.\footnote{Technically the antenna response functions are not frequency independent, however this only becomes a problem when the wavelength of the gravitational wave becomes comparable to the detector arm length.} 

\subsection{Modelled searches}
A comprehensive overview of how GW searches are performed by the LVK collaboration, including how data is calibrated, terrestrial noise vetoed, and how searches and approaches to the construction of GW catalogs have changed over time can be found in \cite{abac2025gwtc}. In this review, we will briefly discuss the principles of GW detection and parameter estimation. 

The detection of gravitational wave signals using matched filtering against a known template waveform is based on a frequentist approach to hypothesis testing \citep{finn92}. To detect gravitational wave signals, we must decide whether the available evidence favors the the null hypothesis\footnote{We have used $H_\mathrm{N}$ to denote the null hypothesis to distinguish it from $H_0$, the Hubble constant, discussed later in this review.}, 
\begin{align}
    \mathcal{H}_\mathrm{N}: d(t) = n(t),
\end{align}
i.e. $d(t) = n(t)$, or whether there is also a GW signal $h(t, \bm{\theta})$ described by parameters $\bm{\theta}$ present in the data stream
\begin{equation}
    \mathcal{H}_\mathrm{S}: d(t) = h(t, \bm{\theta}) + n(t).
\end{equation}
In this case, the likelihood ratio test statistic is equivalent to the matched filter output between the detector data $d(t)$ and a filter $h$ that maximizes the signal-to-noise ratio (SNR, typically denoted $\rho$) \cite{finn92}. It can be shown \citep[e.g. ][]{Findchirp_allen} that the complex matched-filter output between the template $h$ and data $d(t)$ at time $t$ is
\begin{equation}
    z(t) = 4\int_0^{\infty}\frac{\tilde{d}(f)\tilde h^*(f)}{S_n(|f|)}\exp({2i\pi f t})df.
\end{equation}
We normalize the filter output $z$ using the variance of the template 
\begin{equation}
    \sigma^2 = 4\int_0^{\infty}\frac{|\tilde{h}(f)|^2}{S_n(|f|)}df
\end{equation}
which accounts for the fact that the sensitivity of the instrument to a given template signal $h$ is not uniform. We can then write the real-valued SNR timeseries $\rho(t)$ as the absolute value of the two-phase matched-filter output  
\begin{equation}
    \rho(t)= \frac{|z(t)|}{\sigma}.
\end{equation}
which has the appealing property that $\rho^2$ follows a $\chi^2$-distribution with two degrees of freedom \cite{Allen2005} and therefore it is trivial to compute the probability of finding any SNR value greater than some threshold $\rho_\mathrm{thresh}$ as 
\begin{equation}\label{eq:stat}
    P(\rho^2>\rho_\mathrm{thresh}^2) = e^{-\rho_\mathrm{thresh}^2}/2.
\end{equation}
while the expectation value of $\rho^2$ in the presence of a gravitational wave signal is 
\begin{equation}
    \mathbb{E}(\rho^2) = \frac{\mathbb{E}(|z|^2)}{\sigma^2}.
\end{equation}
In practice, a matched-filter search for signals using a single template waveform $h(t)$ discretises the computation of the matched-filter output, and then identifies the time (and phase, for the two-phase filter) at which this output is locally maximized \citep[e.g.][]{cutler94, Findchirp_allen}. To efficiently search for gravitational waves from CBCs, millions of templates for waveforms computed for merging CBCs with masses $m_i\sim 1-400\,\mathrm{M_\odot}$ must be used \citep[e.g. as discussed in ][]{GWTC4_meth}. Template banks are designed by placing the smallest number of templates into a geometric grid of component masses that still maximises the overlap between putative signals. A number of algorithms have been developed to do this \citep[e.g.][]{Allen2021_optimalbanks, gstlalbank2018}, however as the parameter space increases in complexity, stochastic bank generation has emerged as the most efficient way to develop template banks \citep[e.g.][]{Stochastic_banks, Kacanja2024}, especially for searches that incorporate higher-order effects such as eccentricity, spin precession and higher order modes. 

The SNR timeseries' that are produced by filtering GW detector data against this template bank can then be searched for `foreground triggers' that satisfy some threshold criteria for being `signal-like'. Typically, this requires time coincidence between foreground triggers that exceed a limiting SNR threshold: The GW must arrive within the light travel time between detectors (although most searches can identify GW signals present in only one detector data stream, albeit at lower sensitivity). To reduce the impact of non-stationary noise, foreground trigggers must also satisfy some goodness-of-fit criterion between template and putative signal, such as that introduced in \cite{Allen2005}.

Each foreground trigger that satisfies these criteria (typically $\sim 1\,\mathrm{s^{-1}}$), is then ranked against a background noise distribution. In theory, one can compute the false alarm probability of any signal with $\rho>\rho_\mathrm{thresh}$ using Equation \ref{eq:stat}. However, this assumes that the background noise in the detector is purely Gaussian. We know that this is not the case: So-called `glitches' in the detector strain \citep{Davis2022} can be caused by everything from short fluctuations in the laser systems or control systems of the interferometer, to seismic disturbances, lightning or animal activity, to human activity at the detector sites \cite{glitch-rate}. These glitches create triggers that accumulate as heavy tails in the background distribution of $\rho$ measured by the GW detection pipelines, degrading sensitivity to weak signals under the assumption that the detector noise is purely Gaussian. To mitigate this effect, the background distribution of the detection statistic $\rho$ is calculated empirically from detector data streams where coherence and coincidence information is destroyed by performing `time slides' \citep{Was:2009vh} of one detector data stream relative to another. This data is then used to obtain a representative sample of the `background' against which candidates from a coherent search can then be ranked to generate a false alarm probability (FAP). This is converted to a false alarm rate (FAR) by dividing by the total observation time it takes to acquire the background. In general, the FAR scales with the SNR as \citep{CabournDavies2022}
\begin{equation}
    \mathrm{FAR}\propto 10^{-\zeta\rho}
\end{equation}
 where the constant $\zeta\sim2$ and the constant of proportionality are deduced based on the observed detector noise trigger rate. Foreground triggers that satisfy specific criteria, including passing a threshold $\mathrm{FAR}$ are considered GW candidates, and are further scrutinised to evaluate whether they are consistent with a CBC event. 

\subsection{Unmodelled searches}
Matched filtering allows us to search gravitational wave detector data for \textit{known unknowns}: that is, we know we are searching for coalescing compact binary systems, we merely do not know detailed physical information about these systems. However, there are many sources of gravitational waves that cannot be so comprehensively modelled that are still of substantial interest. As noted above, any mass distribution with a non-zero time-varying quadrupole moment can emit gravitational waves that can, in principle, be detected by ground-based laser interferometer experiments provided their frequencies lie within the sensitive band of the detector and their amplitude is sufficiently large. Although unmodelled searches such as \cite{cWB_pipe, BayesWave} are generally less sensitive than optimal matched-filter based searches, unmodelled search algorithms have the benefit of being sensitive to signals that are not well represented in matched filter template banks. These include phenonena such as GWs produced by eccentric binaries and signals with a substantial higher-order mode content, as well as CCSNe and other short-duration burst sources \cite{DAWES_burst}.


\subsection{Single-event parameter estimation}
The detection of gravitational wave transients - whether from modelled or unmodelled searches - represents only one part of a closely-coupled problem. While GW searches may give a rough indication of the nature and properties of the gravitational wave source, more refined inference is required to probe what sort of astrophysical (or more exotic) phenomenon produced the ripples in spacetime. Provided we have a parametric physical model of the GW source, it is possible to estimate the parameters of that source. The most established method to do so is through Bayesian inference. A detailed discussion of the evolution of GW parameter estimation can be found in \citep{Christiansen2022}.

Gravitational wave emission from (quasi-circular, non-precessing, non-tidally deformable) CBCs that can be described by a complete set of 15 parameters $\bm{\theta}$ related to the physical properties of the system \citep{veitch15,ashton19, romeroshaw20}. This framework can be extended to any GW source for which we have a parametric model $h(t, \bm{\theta})$, for example signals from post-merger neutron stars and CCNSe \citep{DAWES_burst}. For a signal $h(t, \bm{\theta})$ in Gaussian noise, the log-likelihood can be written 
\begin{equation}
    \ln\likeli{s_i}{\theta_i}\propto\left<s_i-h(\theta_i),\,s_i-h(\theta_i)\right>
\end{equation}
where 
\begin{align}
    \left<a,b\right>=4{\rm Re}\int_0^\infty df\frac{\tilde{a}(f)\tilde{b}^\star(f)}{S_h(f)}.
\end{align}
is the inner product and $S_n(|f|)$ is the noise power spectral density defined above. This likelihood is called the Whittle likelihood. When combining multiple observatories, $\likeli{s_i}{\theta_i}\propto\prod_j\likeli{s_i^j}{\theta_i}$. That is, in the log-likelihood, the combined multi-observatory likelihood is the sum of each individual observatory log-likehood under the reliable assumption noise in each observatory is uncorrellated, see Fig. \ref{fig:likedag}. The Whittle likelihood is generic to any problem where a well-modelled signal $h(t)$ is superimposed with additive Gaussian noise. This expression should already seem familiar from our discussion of the construction of the optimal filter for GW detection: the SNR is proportional to the square root of the profile likelihood ratio that can be derived from the Whittle likelihood. 

The problem of parameter estimation recasts the frequentist detection problem \textit{`is there any segment of data that is not consistent with the noise hypothesis given our point estimate for the parameter values I am assuming to generate the waveform I am filtering with?'}. Bayes theorem allows us to construct posterior probability distributions on each parameter $\bm{\theta}$ of the waveform model, allowing us to quantify our belief in the value of each parameter, as opposed to a single point estimate. This requires us to have some reasonable expectation on the prior probability distribution of each parameter $\pi(\theta)$ (which can be agnostic or informed by astrophysics). We also need an understanding of the data generation process, which is encapsulated in the likelihood function. The likelihood tells us the probability of measuring data $d$ given the parameters $\bm{\theta}$. We can then use Bayes theorem to construct the 'posterior' probability distribution on the parameters $\bm{\theta}$
\begin{equation}\label{eq:likeli}
    P(\theta|d) \propto \likeli{d}{\theta}\pi(\theta).
\end{equation}
The constant of proportionality missing from this equation is referred to as the evidence $\evid$, 
\begin{equation}
    \evid = \int_\theta \likeli{d}{\theta}\pi(\theta)d\theta
\end{equation}
This corresponds to a normalisation term, but it also plays an important role in hypothesis testing in gravitational wave astronomy. To select between different waveform models that incorporate effects like spin precession or higher order modes, for example, one can construct an odds ratio that incorporates the evidence $\evid$ as a measure of whether the data `prefers' one model over another according to the information content. Generally, the evidence is computationally expensive to compute as it requires one to numerically integrate over a high-dimensional parameter space. However, the structure of the gravitational wave likelihood lends itself well to the use of nested sampling \citep{Skilling2004} algorithms to estimate the posterior probability distribution, calculating the evidence `for free' along the way. 

\subsection{Low-latency detection of gravitational waves}
Gravitational wave detector data can be analysed for the presence of astrophysical signals in close-to real time. Development of these searches has been strongly motivated by interest in the coincident detection of GW emission and electromagnetic or neutrino emission from merging compact objects. A number of gravitational wave detection pipelines search data in real-time, including
\begin{itemize}
    \item Modelled search pipelines PyCBC Live \citep{PyCBC, Davies2020}, GstLAL \cite{gstlal_paper,Sachdev:2019vvd}, MBTA \citep{MBTA, MBTA_pipeline}, and SPIIR \cite{spiir_shaun, SPIIRO3},
    \item Unmodelled search pipeline cWB \citep{cWB_pipe},
    \item Weakly-modelled searches incorporating machine learning techniques: Aframe \citep{AFrame} and cWB-BBH \citep{cWB_pipe, cwb-xgboost} (and references in \citep{abac2025gwtc}).
\end{itemize}
These pipelines analyse data acquired by the LIGO, Virgo and KAGRA interferometers in real-time, and have done so since the beginning of the advanced detector era\footnote{Although the Advanced Virgo interferometer joined joint operations only midway through O2, with KAGRA joining intermittently during O3 and O4.}. Since the beginning of the third observing run (O3), alerts were released to the general public.

Although GW detection algorithms can operate with little to no time delay between data acquisition and analysis due to low computational overheads, until recent years the parameter estimation process has operated at much higher latencies, typically on the order of several hours, to several days \citep{LIGOdata_rev}. This, coupled with interest in following up GW detections with electromagnetic telescopes with limited fields of view, has motivated the development of rapid parameter estimation algorithms. BAYESTAR \citep{bayestar} is used to compute localization of CBC sources in low-latency, producing skymaps that can be ingested by followup campaigns. Over time, software developments have reduced both detection and parameter estimation latencies \citep{GWTC4_meth}. For example, the implementation of algorithms on GPUs \citep{xiaoyang2018, Bilby_gpu} has provided substantial speed-up of both matched filtering and parameter estimation. A variety of other data processing techniques have also shaved off critical fractions of a second in latency, including time-domain data whitening \citep{Tsukada2018}. In the context of parameter estimation, latencies have been substantially reduced thanks to the advent of massively parallel implementations of nested sampling and the development of new frameworks to assist in the parameter estimation workflow \citep{pbilby, Rift, Rift2, Asimov}. Machine learning has also been used to great effect in GW detection and parameter estimation, motivated primarily by latency reduction \citep[see][ for a review]{CuocoLRR}.

As discussed above, the primary heuristic that is used to decide whether or not a candidate identified by a GW detection pipeline is a `real' GW signal or not is the false alarm rate (FAR). Calculation of the FAR is based on the ranking of `foreground triggers' against a noise background and contains little to no information about the known or expected population of astrophysical gravitational wave sources. There is implicit information in the sense that modelled searches are based on a template-bank that has been chosen based on some expectation about the population, however this does not explicitly propagate to the hypothesis being tested in the formation of the detection statistic.  A threshold FAR of $\mathrm{FAR}<1\times10^{-8}\,\mathrm{s}$ was chosen to filter candidates that were of sufficient significance to trigger futher investigation and dissemination of the GW alert to the public \citep{GWTC3}. In O4 candidates with $1\times10^{-4}\,\mathrm{s^{-1}}< \mathrm{FAR}<1\times10^{-8}\,\mathrm{s^{-1}}$ were reported as subthreshold candidates, with information about these candidates also released publically \citep{GWTC4}. 

The FAR only quantifies the rate at which one would expect to see a similar trigger arise from noise processes within the detector, and it does not give any estimation of how likely it is that the trigger is of astrophysical origin. The latter is of great interest to the multimessenger astronomy community. To determine the \textit{probability of astrophysical origin}, $p_\mathrm{astro}$, requires knowledge of the source population as well as the sensitivity of search algorithms. $p_\mathrm{astro}$ accounts for the true positive rate, and is more consistent when working with catalogs containing events that have been detected with different selection functions \citep{Banagiri2023}. The formal mathematical definition of $p_\mathrm{astro}$ is \citep{GWTC4_meth}
\begin{equation}
    p_\mathrm{astro}(r) = \frac{\mathcal{R}_af(r)}{\mathcal{R}_af(r) + \mathcal{R}_nb(x)} 
\end{equation}
where $\mathcal{R}_a$ ($\mathcal{R}_n$) is the total rate of signal (noise) candidates identified by a pipeline, and $f(r)$ ($b(r)$) are the probability density function of the signal (noise) events at ranking statistic $r$. There are differences in how $p_\mathrm{astro}$ is calculated by each pipeline: $f(r)$ is based on an astrophysical source model that differs slightly between pipelines \citep[see][for details]{GWTC4_meth}, and each pipeline uses its own search statistics to estimate the rates $\mathcal{R}_x$ that depend on the method used to calculate the FAR of candidates. However, despite methodological differences, $p_\mathrm{astro}$ values are broadly consistent across pipelines \citep{Sharmachoudary2024}. $p_\mathrm{astro}$ was also one of the metrics released through open public alerts since the beginning of the third LVK observing run, and in LVK public alerts is broken down into $p_\mathrm{astro} = p_\mathrm{BBH} + p_\mathrm{NSBH}+p_\mathrm{BNS}$. These classes are based on a somewhat arbitrary mass cut-off to delineate NS and BH ($M_\mathrm{BH|NS}=3\,\mathrm{M_\odot}$), and $p_\mathrm{astro}$ contains no information about the EOS \citep{Sharmachoudary2024, IGWN_OPA}. The reciprocal of $p_\mathrm{astro}$ is $p_\mathrm{terr}$, i.e. the probability that the trigger resulted from non-stationary noise in the detector as a result of terrestrial activity. High $p_\mathrm{terr}$, together with clear evidence of terrestrial noise, occasionally results in the retraction of GW candidate alerts.

Over time, additional heuristics have been introduced (all calculated on the assumption of an event being of astrophysical origin) to inform electromagnetic follow-up of GW events. These metrics include:
\begin{itemize}
    \item \textit{EM-Bright}: CBCs containing at least one neutron star have the potential to produce electromagnetic emission through the tidal disruption and heating of neutron-rich material \citep{metzger10}. Such a signal is dependent on the masses and equation-of-state of the compact objects that merge. \texttt{EM-bright} \citep{Chatterjee_2020} provides machine-learning tools to compute three scores that can be used to gauge the likelihood a given gravitational wave event: contains at least one neutron star (\texttt{HasNS}), leaves behind tidally-disrupted matter outside the remnant compact object (\texttt{HasRemnant}, which is most pertinant to NSBH systems), and produces a remnant compact object in the lower mass gap (\texttt{HasMassGap}). \texttt{EM-bright} marginalises over equation-of-state, and mitigate\jm{s} the potential systematic and random errors that can arise from triggers produced by template based searches. During O4, an additional category was introduced to account for the introduction of sub-solar mass (SSM) searches (\texttt{HasSSM})\footnote{Currently, no $p_\mathrm{astro}$ is calculated for SSM searches, \citep{IGWN_OPA}}.
    \item \textit{Binned source chirp masses}: Toward the end of O4, the LVK began to release binned source-frame chirp mass estimates for both modelled search pipelines (which produce a coarse maximum likelihood estimate of the detector-frame chirp mass and the source redshift through the matched filtering process) and for the weakly-modelled cWB-BBH search. Information is released in the form of the probability that the source-frame chirp mass belongs to a coarse mass bin \citep{IGWN_OPA}. 
\end{itemize}

\subsection{Public data releases}
GW detector data can be obtained from the Gravitational Wave Open Science Center \citep[GWOSC,][]{GWOSC}, making it possible for searches for gravitational wave transients to be done by scientists external to the LVK collaboration. For example, \cite{Zackay2021} identify a binary black hole merger not identified in GWTC-1 \citep{GWTC1}. Their search improves on the sensitivty of the LVK searches by accounting for non-Gaussian fluctuations in the detector noise. The Open Gravitational Wave Catalogs \citep[OGC, ][]{1OGC, Nitz2020, 3OGC, 4OGC} also analyse data from LVK observing runs. Although 1-OGC \citep{1OGC} does not identify any additional candidates beyond those reported in GWTC-1, 2-OGC \cite{Nitz2020} identifies GW151205, a BBH merger with a primary mass of $67^{+28}_{-17}M_\odot$. The OGC search uses the PyCBC matched filter gravitational wave detection pipeline framework, but incorporates a different algorithm to select graivtational wave signal candidates by imposing phase, amplitude and time delay consistency \cite{1OGC}. The significance of the candidates is calculated using an improved background model, and a larger search parameter space of signal templates. These searches are important: they validate the methods used to construct GWTC-1 and demonstrate that the results are robust to differences in detection methods. They also show the power of the wider astronomy community being able to access the complete LVK data sets that are used to produce the GWTC and the value of open data.

\subsection{Gravitational wave discovery in action: GW150914}
On September 14, 2015 the two detectors of LIGO simultaneously observed their first gravitational wave transient: GW150914 \cite{gw150914}. The event was initially identified by the cWB pipeline \citep{cWB_pipe}, an unmodelled search for generic gravitational wave transients, which issued an alert within three minutes. Within a few hours, additional searches for generic gravitational wave transients, omicron-LALinference-Bursts (oLIB, \cite{olib}) and BayesWave \cite{BayesWave}, confirmed the presence of a signal. A matched-filter search of the GW data obtained between September 12 and October 20 2015 was subsequently carried out using the PyCBC-Live and GstLAL searches. Both of these pipelines require stable instrument operation over periods longer than $\sim 1\,\mathrm{week}$ to generate reliable estimates of event significance \cite{gw150914}. Given that GW150914 was detected just two days after the detectors were switched on, it is therefore not surprising that these pipelines did not detect the event in low-latency. GW150914 was identified with the same matched-filter template in both analyses, with component masses $47.9\,\mathrm{M_\odot}$ and $36.6M_\odot$ at a combined matched-filter SNR $\rho_c = 24$. The two search algorithms perform ranking of GW candidate triggers differently, however both resolve consistent false alarm probabilities: from PyCBC, only a upper limit on the FAP can be derived, as $\mathrm{FAP}<2\times10^{-7}$ over 384 hours of observation, whereas from GstLAL, the $\mathrm{FAP} = 1.4\times10^{-11}$ \cite{gw150914}.

\subsubsection{What did GW150914 tell us about astrophysics}
From an astrophysical perspective, the detection of GW150914 revealed that not only can black holes with masses $25M_\odot$ form in binary systems in nature, they can merge within a Hubble time \cite{abbott2016astrophysical}. The results of full parameter estimation using LALinference for the first GW event detected using laser interferometers revealed a binary of chirp mass $\mathcal{M} = 28^{+2}_{-2}M_\odot$, comprising components $m_1 = 36^{+5}_{-4}M_\odot$ and $m_2 = 29^{+4}_{-4}M_\odot$ \citep{abbott2016astrophysical}\footnote{The discrepancy between the masses determined by the detection pipelines and parameter estimation is explained in Sections 3.1 and 3.3}. The system was revealed to have an effective spin consistent with $\chi_\mathrm{eff}\sim0$. The formation of black holes with masses consistent with the components of GW150914 through binary stellar evolution requires weak massive-star winds, which are expected to arise only in systems with comparatively low metallicities $[Z/Z_\odot]\le 0.5$ \citep[e.g. ][]{Eldridge2016, Andrews2021}. On the other hand, the observed masses can also be explained through dynamical interactions leading to the formation of the BBH system that merged to produce GW150914 \citep{rodriguez2016dynamical}.

\section{Statistical properties of GW source populations as a messenger}\label{sec:pops}
Fundamental questions about the astrophysics of compact objects can only be answered by considering the statistical properties of the overall CO population. These questions include whether there is a limit on the maximum mass of a black hole that forms through stellar collapse, and what the overall structure of the black hole and neutron star mass distributions tell us about star formation, stellar evolution across cosmic time. We can also probe their formation mechanisms and evolutionary history. Each proposed channel for forming a compact binary — isolated binary stellar evolution, dynamical assembly in globular or nuclear star clusters, evolution through active galactic nuclei discs, or hierarchical triple systems — makes testable predictions about the resulting distributions of masses, spins, mass ratios, and merger rates \citep[see ][and references therein, for example]{gerosa2021hierarchical}. By constructing flexible and physics-informed statistical models for these distributions and fitting them to the ensemble of detected events using hierarchical Bayesian inference, it is now possible to confront these predictions with data.

\subsection{From single-event to population inference}
\begin{figure}
\centering
\includegraphics[width=\columnwidth]{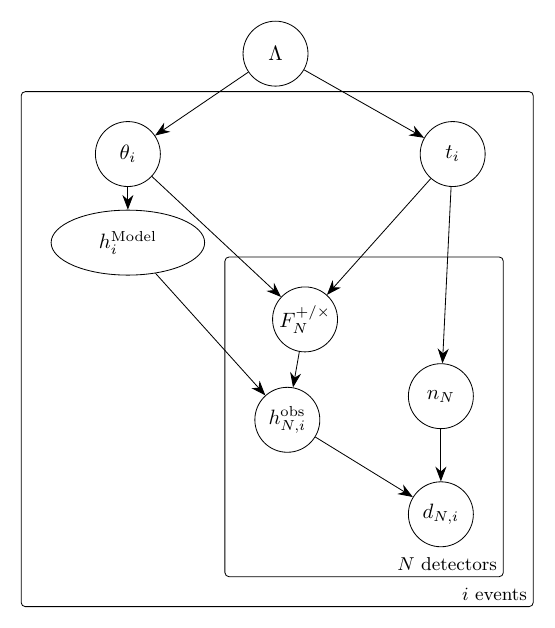}
\caption{Directed acyclic graph showing the hierarchical strucutre of the GW population inference problem and the interdependence of the components of the population likelihood Eq. \ref{eq:poplike}. Based on Fig 1 of \citep{Essick2025}}\label{fig:likedag}
\end{figure}

Beyond single-event inference, which provides insight into the properties of individual events, it is also important to determine the properties of the GW source population based on catalogs of (sometimes uncertain) observations. The hierarchical Bayesian inference formalism used to do this is able to consistently account for measurement uncertainty and selection effects to infer `population hyperparameters' that describes a model from which individual event parameters are drawn. The population likelihood used in GW population inference is the joint likelihood of the dataset $d$ comprised of $d_i$ observations given a population model $M(\bm{\Lambda})$ that is described by the population hyperparameters\footnote{Not to be confused with the Cosmological Constant, which is also denoted $\Lambda$, but serves as short-hand for a negative-pressure equation of state rather than an inferrable parameter} $\bm{\Lambda}$,
\begin{equation}\label{eq:poplike}
    \likeli{d}{\bm{\Lambda}, M} = \prod_{i=1}^{N}\frac{1}{\xi(\bm{\Lambda})}\int\likeli{d_i}{\theta_i}\pi(\theta_i|\bm{\Lambda})d\theta_i
\end{equation}
The likelihood is marginalized over the parameters of each individual observation $\theta_i$, and $\pi(\theta_i|\bm{\Lambda})$ is a conditional prior that defines the population model $M$. This is equivalent to the population likelihood after marginalizing over the scale of the Poisson process that describes the GW event detection rate given a specific choice of prior (for example, see the derivation of \cite{Mandel2019}, which gives a bottom-up construction of the population likelihood). The relationship between the population likelihood and the individual event likelhood can be visualized as a directed acyclic graph (Fig. \ref{fig:likedag}) that shows how the individual event likelhood components, namely the waveform model for the i-th event $h^\mathrm{Model}_i$ and its parameters $\theta_i$ are drawn from the population hyperparameters $\Lambda$. The likelihood in Eq. \ref{eq:likeli} represents the process that generates the data observed by each of the $N$ detectors in the inner panel. 

The $\xi(\Lambda)$ normalization factor is used to account for the observational selection function associated with GW searches \citep{cutler94}. Our observed population of CBCs is the result of a censored process, where only some subset of the population meet the criteria for detection. The selection function responsible for this can be written 
\begin{equation}
    \xi(\Lambda) = \int P_\mathrm{det}(\theta)\pi(\theta|\Lambda, M)d\theta.
\end{equation}
Determination of $P_\mathrm{det}$ is non-trivial and is generally evaluated through large scale `injection campaigns' where known signals drawn from a reference population are injected into detector noise, and search pipelines investigate the rate at which they can recover these signals at a given significance threshold \citep[e.g.][]{Tiwari_2018, Farr_2019, Essick_2021, Essick2025}. This allows us to quantify the relative fraction of events in our astrophysical population we would expect to 'miss' given standard search strategies as a function of the GW signal parameters. The introduction of the population likelihood allows us to access an additional `messenger' -- the population hyperparameters $\bm{\Lambda}$ --  from which we can draw astrophysical conclusions. 

\subsection{Binary black hole population}
Early attempts to resolve structure in the BBH mass distribution rely on comparatively simple models that are not over-specified, given the small size of the CBC population detected during O1 and O2. For example, in \cite{BBHdist_LVK1} the only free parameter for the population model is the spectral index of the power-law distribution of the primary mass $m_1$, 
\begin{equation}
    p(m_1)\propto m_1^{-\alpha},
\end{equation}
which was inferred to be $\alpha = 2.5^{+1.5}_{-1.6}$ from five BBH events. This is consistent with a flat distribution of $m_1$. Simple models allow for easy interpretation of the effects of model misspecification: the flat distribution is driven by a high, fixed upper limit on the maximum black hole mass. This was broadly consistent with expectations from different population synthesis simulations \cite{Abadie2010} as well as dynamical simulations, and allowed early constraints to be placed on the typical metallicity of the environments in which the five BBH merger systems must have formed. The shape of the mass distribution was more tightly constrained as the modelling increased in complexity and more data became available with \citep{GWTC1}, where population models incorporate parameterisations of the mass distribution, spin distribution, and merger rate, whose population parameters (hyperparameters) are inferred jointly from the data \citep[e.g. following formalism developed in ][]{Talbot2017, Fishbach_2018, Wysocki2019}. It became clear that the primary mass distribution is not flat, and in fact falls quite steeply with $\alpha\sim 1.5$. The first hints of structure surfaced, with marginal evidence for a peak in the mass distribution around $m_1\sim 10\,\mathrm{M_\odot}$. Conclusive observational evidence for this peak was first found based on GWTC-2 \citep{Tiwari2021, Edelman2022} and subsequently incorporated into the weakly modelled LVK population analysis in GWTC-3\citep{GWTC3_pop}. CCSNe mechanisms associated with the formation of BHs with masses $\sim 10\,\mathrm{M_\odot}$ are relatively well understood \citep{Burrows_CCSNe_10} and this feature is predicted in isolated binary evolution modelling \citep[e.g. ][]{Dominik2015, Belczynski2020}

Evidence for structure in the mass distribution is not limited to this very prominent peak in the merger rate densities around $10\,\mathrm{M_\odot}$. There is also now evidence that the power-law distribution of $m_1$ is not consistent with a single power-law, and there is instead a break around $30\,\mathrm{M_\odot}$ \citep{gwtc4pop}. Interpretations of this feature are scant, but could point at either a dynamically forming sub-population or a mass-dependent efficiency in the rate at which binary black holes form through binary stellar evolution (e.g. variations in the efficiency of secular processes such as common envelope phases, mass loss, or mass reversals \citep{Belczynski2002}). Structure around $m_1 \sim 35\,\mathrm{M_\odot}$ has been present in each population model since GWTC-2 \citep[e.g. Model C of ][]{GWTC2-pop} however was not identified at the time at high credibility. Interestingly, this feature is well evidenced in the GWTC-4 analysis \citep{gwtc4pop} across both strongly and weakly modelled approaches to fitting the overall population, not just a pathology of the approach taken in \citep{GWTC2-pop}. This population of BBH mergers with masses $\sim 35\,\mathrm{M_\odot}$ preferentially merge with more equal mass partners \citep{Fishbach2020, gwtc4pop}. Peaks in the BBH mass distribution from dynamical simulations of dense stellar cluster are expected to occur at masses $>10\,\mathrm{M_\odot}$ \citep{Rodriguez2016_mergerrates, rodriguez2019black}, and may contribute to these observed sub-populations. 

With the advent of weakly-modelled approaches to population modelling (such as \citep{Mandel:2016prl, Li:2021ukd,Ray:2023upk}, as selected examples), it is tempting to assume that any feature in the mass distribution can be interpreted as an additional feature or evidence for further sub-populations. However, the apparent \textit{absence} of one key feature that is predicted by the majority of stellar endpoint modelling is particularly notable. The pair-instability mass gap \citep{Woosley2021} is likely not empty -- both GW190521 \citep{gw190521} and GW231123 \citep{GW231123} have components that have masses inconsistent with having formed through stellar collapse -- tells us that we do not fully understand the mechanisms of CCSNe or we are finding hierarchical mergers. Coupled with the evidence from BH spin distributions, there is good evidence that at least some of these are forming hierarchically, although a primordial origin of GW231123 has been posited \citep{GW231123_primordial}, and the presence of a mass cut-off is somewhat sensitive to modelling and prior assumptions \citep{Ray:2025xti, Mould2026}. Looking beyond the distribution of primary masses, there is now some emerging evidence that there could be an upper mass-gap in the $m_2$ distribution \citep{Tong2026}.

The mass ratio distribution of the population also encodes important astrophysical information as it can hint at formation mechanisms. The first approaches to explicitly modelling the mass-ratio distribution along with the primary mass distribution found that the mass ratio distribution tends to favour comparable mass binaries, although detection of events such as GW190814 and GW190412 in \citep{GWTC2-pop} indicate that BBH with mass ratios that are substantially non-equal do exist in nature at a non-negligible rate. Now, analysing a population of around 200 events, it is clear that there the aforementioned peak in the distribution around $m_1\sim 10\,\mathrm{M_\odot}$ is a statistically significant feature, and there is a fairly well-established correlation between this peak and mass-ratio $q$, suggesting a sub-population that preferentially pairs up with smaller black holes where $q\sim 0.7$ \citep{Godfrey:2023oxb,Banagiri2025}. 


The spin distribution gives us valuable evidence about the origin of BBH systems, potentially allowing us to distinguish sub-populations arising from isolated binary evolution, and dynamical formation in a variety of different environments. Early constraints on the spin magnitude $\chi$ and $\chi_\mathrm{eff}$ distributions suggested that BHs tend to be born with low spins or some other process causes the spin vector orientation to tilt in such a way that $\chi_\mathrm{eff}$ becomes small \citep{Belczynski2020}. However, when the spin magnitudes are small enough, simple models such as those used in early analysis of BBH spins tend produce features that cannot be distinguished within observational uncertainties \citep{GWTC2-pop}. In \citep{GWTC3_pop} there was a striking claim: 12-44\% of BBH have $\chi_\mathrm{eff}<0$, implying at least one black hole tilt angle is $>90\deg$ (retrograde spins). The same analysis finds evidence for an overdensity of BBHs with $\chi<0.01$. The latter subpopulation of binaries with negligible spin and no evidence for significantly misaligned spin would be consistent with the field formation scenario \citep{Fuller2019, Mandel2022}, while the former, with misaligned spins with large tilt angles could most plausibly form through dynamical interactions \citep{rod18}. 

There is ongoing debate about the existence of this overdensity at small values of $\chi$: on one hand, works find that current data does not require the inclusion of a low-spin subpopulation in the spin distribution model \cite{Callister2022, tong2022} or find that only a small fraction of events posses low spins \cite{Kimball2020}. Others find evidence that is supportive of two distinct sub-populations in the spin distribution\citep{Roulet2021, Galaudage2021}, including a low-spin sub-population \citep{Biscoveanu:2020are, Mould:2022xeu}. Analysis of the GWTC-4 population confirms a dearth of BBH with large spins, with $90\%$ of events having $\chi<0.57$. Both models produce broadly consistent results: a peak in $p(\chi)$ at $\chi \sim 0.01-0.23$ under the assumption that the spins are independently and identically distributed\citep{gwtc4pop}. 

There is also evidence that at least one black hole per binary has $\chi\geq0$ \citep{gwtc4pop}. This finding is particularly interesting in the context of astrophysics as it suggests that at least one BH per binary is spun up by some process in its evolution, for example through tidal interactions, or through accretion \citep{Belczynski2020}. Low numbers of detected BBH with large spin magnitudes suggest that although there is a non-negligible contribution from hierarchical mergers, the population cannot be dominated by second-generation BHs \citep{Fishbach:2022lzq, Vijaykumar_2026}. 

The $\chi_\mathrm{eff}$ distribution in GWTC-4 is found to be skewed and asymmetric with more support for positive values \citep{gwtc4pop}. This observation is consistent with a dominant sub-population arising from isolated binary evolution where spin-orbit alignment is retained throughout the evolution of the binary. There is, however, a non-negligible number of binaries with negative $\chi_\mathrm{eff}$ (fraction of $0.24-0.42$ depending on the spin model used). This is used to bound the fraction of BBHs that form dynamically in gas-free environments (as gas-rich environments are thought to contribute toward spin alignment) under the assumption that the spins are isotropically distributed. This has allowed constraints on the fraction of BBH that form dynamically: less than $84\%$. Of the BBH population that form dynamically, hierarchical mergers are also expected to have negative $\chi_\mathrm{eff}$, and is calculated to be $\leq 3\%$ in the analysis of \citep{gwtc4pop} following \citep{Baibhav2020, Fishbach2022}, although this value is highly model dependent. With the observation of more BBH, measurements of the $\chi_\mathrm{eff}$ distribution will become more informative, however there is increasingly compelling evidence that the skewness of the distribution arises from an astrophysical process rather than model misspecification \citep{Banagiri:2025dxo}.

This, coupled with the information about the spin magnitudes discussed above, suggests that the majority of the observed population cannot be forming through dynamical mergers, and hierarchical mergers contribute a small (but still non-zero) fraction of the observed population \citep{gwtc4pop}. 

\begin{figure}[h]
\centering
\includegraphics[width=\columnwidth]{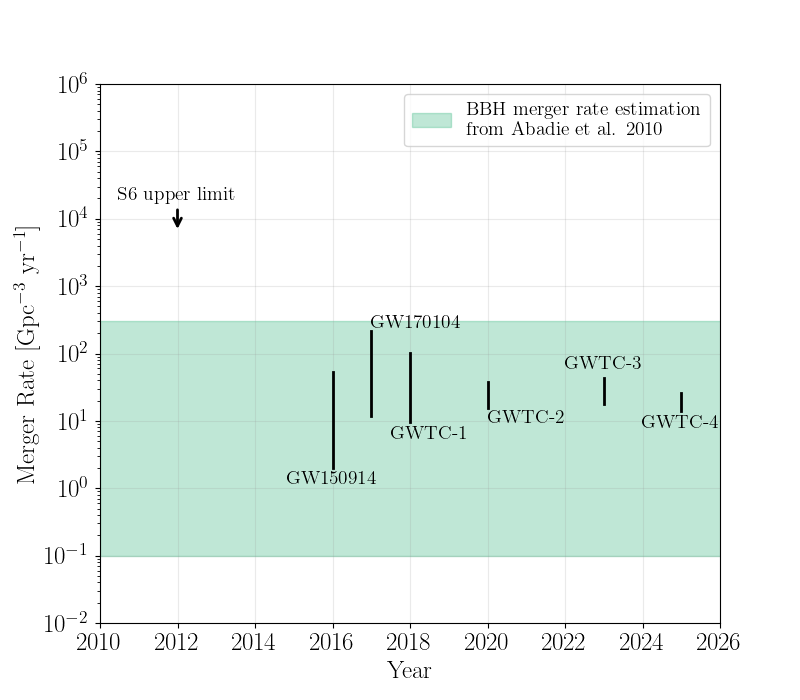}
\caption{Observed GW BBH rates from \citep{Abadie2010, S6LIGO, gw150914, GW170817, GWTC1, GWTC2-pop, GWTC3_pop, gwtc4pop}}\label{fig:bbh}
\end{figure}

Insights into the origins and evolution of BBH can be gleaned from the overall BBH merger rate, and its evolution across cosmic time.  More recent theoretical estimates of the BBH merger rate from stellar evolution simulations are summarised in \cite{Mandel_floor2022}. A comparison of the merger rates predicted by \cite{Abadie2010} and the observed merger rate over the course of the initial and advanced detector observations is shown in Figure \ref{fig:bbh}. Beyond the overall present day merger rate, the delay-time distribution (DDT) is the time between the formation of a star and it's demise as a binary black hole merger and is highly dependent on the processes that shape isolated binary evolution, or the timescales on which dynamical formation of BBH can occur. The delay times of BBH in dynamically formed and isolated binaries can vary from Myr to more than a Hubble time \citep{Neijssel2019, Giacobbo_massloss, Giacobbo_metallicity} depending on phenomena such as common envelope phases, stable mass transfer, and mass-ratio reversal \citep{Mandel2016, Rodriguez2016_mergerrates, Rodriguez_2018}. It is now possible to observationally infer some information about the BBH delay-time distribution \citep{Fishbach2021} indirectly through either studying the rate evolution of BBH mergers as a function of redshift. Observations of the redshift evolution of the BBH population are challenging to confront with theoretical predictions, due to a relative dearth of data at higher redshifts due to the limited horizon of current generation BBH detectors. Theoretical predictions may also overestimate BBH formation rates relative to detailed stellar evolution models \citep{gallegos2021transfer}, indicating that deeper understanding of binary stellar evolution may be required to fully resolve the DTD. Nevertheless, observations currently disfavour very long BBH delay times, suggesting that common envelope phases are extremely important in shaping the observed BBH population that forms in isolated binaries. The influence of dynamical interactions on the delay time distribution is still relatively under-explored, especially in gas-rich environments \cite{Delfavero2025McfactsIII}.

\subsection{NS and NSBH populations}
\begin{figure}[h]
\centering
\includegraphics[width=\columnwidth]{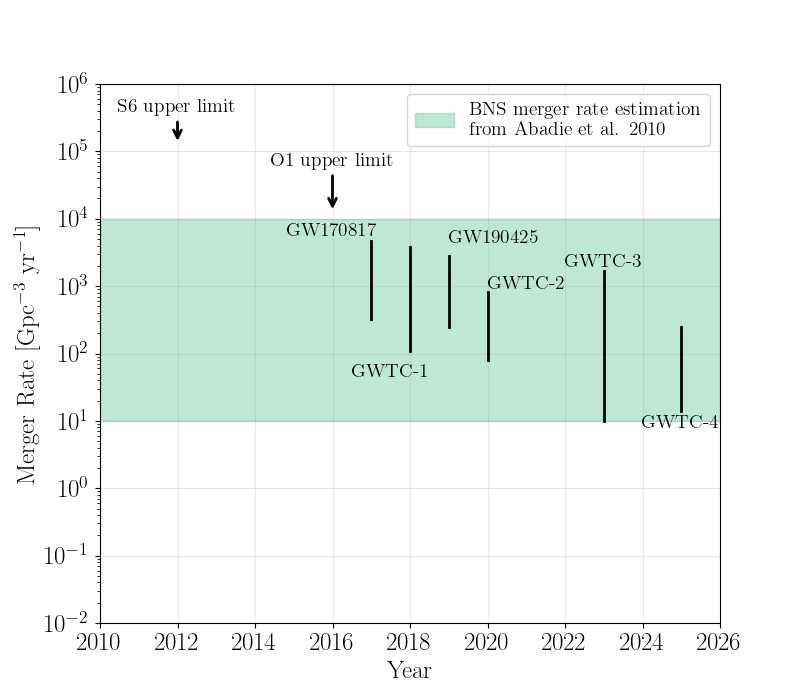}
\caption{Observed GW BNS rates from \citep{Abadie2010, S6LIGO, O1upperlimits, GW170817, GWTC1, GW190425, GWTC2-pop, GWTC3_pop, gwtc4pop}}\label{fig:bns}
\end{figure}

Unlike binary black holes, the existence of neutron star binaries -- and the emission of gravitational waves by neutron star binaries -- was well established by the time of the first direct detection of gravitational waves in 2015. The population of galactic double neutron stars (albeit small) set the expectations for BNS gravitational wave sources, with extrapolations of the BNS merger rate from the galactic DNS population and population synthesis suggesting merger rates of between 0.4 - 400 per year (90\% CI) \citep[][and references therein]{Abadie2010} (Fig \ref{fig:bns}). By the beginning of the second LVK observing run in 2017, it seemed that a detection of a binary neutron star merger was only a matter of time.

Until 2017, everything we knew about the mass distribution of BNS systems came from electromagnetic observations of both isolated and binary Galactic pulsars. Analysis of the Galactic pulsar population shows a double-peaked mass distribution \citep{ozel2012mass, Kiziltan2013, Farr2020_NS, Shao2020}. The secondary peak lies close to $M\sim1.7\,\mathrm{M_\odot}$. Only isolated pulsars are found in this higher mass component: the Galactic \textit{double} neutron star population appears to consist exclusively consist of sources with masses $m_{1,2}\sim 1.35\,\mathrm{M_\odot}$. There are theoretical constraints that bound the range of permissible neutron star masses: causality limits the maximum mass of a stable neutron star to $<3\,\mathrm{M_\odot}$ \citep{Kalogera1996} while a lower bound of $1\,\mathrm{M_\odot}$ is imposed by the dynamics of plausible core collapse supernova mechanisms \citep[e.g.][ and references therein]{Muller2025}. Indeed, there are no galactic neutron stars that have been identified outside this mass range. A more stringent limit on the maximum neutron star mass is the Tolman-Oppenheimer-Volkoff maximum mass for a non-rotating neutron star \citep{Tolman39, Oppenheimer39}. The precise value of $M_\mathrm{TOV}$ is strongly dependent on the (currently unknown) equation-of-state (EoS) but expected to be around $M_\mathrm{TOV}\sim 2.0-2.2 M_\odot$ based on observations of galactic neutron stars \citep{Alsing2018}. The most massive pulsars detected in the Galactic population are thought to have masses that lie close to the TOV mass, allowing the exclusion of some extreme EoS \citep{Shao2020}.

Gravitational waves offer a new avenue to explore the neutron star mass distribution. The first BNS merger GW170817 revealed that this extra-galactic DNS system had properties broadly consistent with the known Galactic DNS population \citep{GW170817_props}. However, the discovery of GW190425 \citep{GW190425} revealed that merging BNS systems that contain far more massive NS do exist in nature, and that the merging DNS population may differ from the DNS population in our galaxy.

Electromagnetic observations have not yet detected binary systems consisting of black holes and neutron stars. Identifying such a system, for example comprised of a stellar mass BH and a pulsar is a key objective for radio astronomy experiments, and such systems have been conjectured to exist for based on simulations \citep{Debatri2021} and observations of high-mass x-ray binary systems \citep{ChatyNSBH}. Through O1 and O2, it was only possible to place an upper limit on the rate of NSBH mergers of $\mathcal{R}\leq610\,\mathrm{Gpc^{-3}\,yr^{-1}}$ \citep{GWTC1}. This upper limit is, nevertheless, weakly constraining of predictions from binary population synthesis estimates which are highly sensitive to the complex processes that occur during the evolution of massive binary stars and range from $0.1-800\,\mathrm{Gpc^{-1}\,yr^{-1}}$ \citep{Abadie2010}.

\begin{figure}[h]
\centering
\includegraphics[width=\columnwidth]{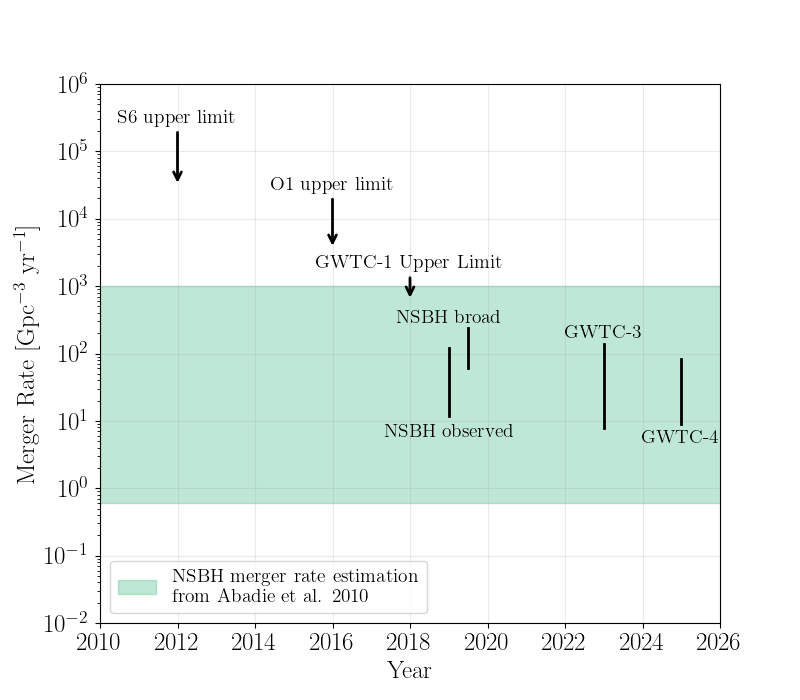}
\caption{Observed GW NSBH rates, from \citep{Abadie2010, O1upperlimits, GWTC1, NSBHdisco, GWTC3_pop, gwtc4pop}}\label{fig:nsbh}
\end{figure}

Our first indication that NSBH systems do exist in nature came toward the end of the Third LVK observing run \citep{GWTC3}. Two NSBH events were detected in quick succession: GW200105 ($9+1.9\,\mathrm{M_\odot}$) and GW200115 ($6+1.5\,\mathrm{M_\odot}$) \citep{NSBHdisco}. Although GW200105 was later found to not pass the $p_\mathrm{astro}$ threshold for inclusion in GWTC-3 \citep{GWTC3}, it has remained a candidate of interest. GWTC-4 \citep{gwtc4pop} confirmed only one NS-containing system with $\rm{FAR}<0.25$ yr for inclusion in their population analysis, namely GW230529 \citep{GW230529} (although a second was identified in the engineering run preceeding O4a, GW230528, which is excluded from the population analysis as its detection involved substantial human intervention). 

In spite of the small number of NS-contaning CBCs that have been detected, the mass distribution of neutron stars can still be inferred. GWTC-3 found that the inferred NS mass distribution was relatively flat \citep{GWTC3_pop}, with a sharp decrease in the merger rate at $m\sim 2.5\,\mathrm{M_\odot}$. In GWTC-4, one additional NS (the secondary component of GW230529) is added to the population \citep{gwtc4pop}\footnote{There are more potential NS in GWTC-4, however only GW230529 passes the threshold to be included in the high-purity population analysis.}. Two different models are used to fit the NS mass distribution: one in which masses are assumed to be Gaussian distributed, and another in which they are assumed to follow a power law.  Analysis shows that neutron stars detected through GW emission follow a mass distribution with a power-law slope constrained to $\alpha = 7.7$, or in the case of the Gaussian model  shows largely unconstrained peak width $\sigma = 0.68+1.2$ and location $\mu = 1.4+0.48$. Although these results are highly uncertain due to the small population size, there is evidence for a peak in the mass distribution that is consistent with the Galactic DNS mass distribution peak around $1.3\,M_\odot$, however the distribution of NS masses in GW-emitting systems is much broader.

As discussed above, LVK analyses make stringent cuts on the FAR and $p_\mathrm{astro}$ of events that are incorporated into population analyses. This ensures a high level of purity (low risk of incorporating noise transients) and data that is of sufficient quality to extract significant amounts of information. In the context of the NS mass distribution, this also involves excluding events where there is ambiguity about whether a CBC component is a BH or NS. This results in the exclusion some events that have been interpreted as containing NS, such as GW190814. This event has been variously classified as a BBH or NSBH merger \citep{GWTC3_pop, Read2021}. The inclusion of the secondary component of GW190814 \citep{gw190814} tends to flatten the NS mass distribution and extend it beyond $2.5\,\mathrm{M_\odot}$ \citep{Read2021}, potentially providing evidence that the GW BNS population differs from the Galactic population \citep{Galaudage2021_NS}. The more parsimonious interpretation is, however, that GW190814 is a BBH merger \citep{Essick2020_NSmass} 

GW190814 and events like it present a classification problem. Prior to the detection of gravitational waves, electromagnetic observations have suggested there is a dearth of compact objects with masses between $3-5\,\mathrm{M_\odot}$, a region occasionally referred to as the `lower mass gap'. The processes that are thought to occur during the core-collapse of massive stars, such as neutrino heating, can be invoked to explain the presence of this apparent separation between the black hole and neutron star mass distributions. However, we are now detecting more and more objects - in both EM and GW observations - that suggest the lower mass gap may not be empty at all. For example, the primary component of the merging system that produced GW230529\citep{GW230529} has posterior support for $m_1$ inside the lower mass gap. The interpretation of this events as BNS or NSBH is uncertain and depends on the assumed mass and spin priors. A clear detection of tidal deformability induced phasing of the gravitational waveform could be considered evidence for the presence of at least one neutron star component, however this requires signals with high SNR.

Gravitational wave detection treats classification of events as BNS, NSBH or BBH as a comparatively simple problem based on arbitrary definitions of allowed component masses of various systems (see discussion of $p_\mathrm{astro}$ above). In GWTC-3 \citep{GWTC3_pop}, NS and BH categories were distinguished using prior information on the maximum neutron star mass based on constraints on $M_\mathrm{TOV}$ that constrain the maximum non-rotating neutron star mass $M_{TOV}<2.5$. \citep{GWTC3_pop} A similar method was used in the analysis of GWTC-4 \citep{GWTC2-pop}, which explicitly excluded GW190814 from the NS mass distribution analysis based on its ambiguous classification. More rigorous statistical methods of handling this classification problem such as \cite{Essick2020_NSmass} have been proposed, but a simplistic sub-division of the population based on heuristic arguments about the TOV mass has been preferred.


\subsection{Cosmology as population-level inference}
Because each gravitational wave signal detected carries with it some information about the location of its source in the cosmos, we can use compact binary mergers as `standard sirens' \citep{hughes05} to study the expansion history and energy density content of the Universe. The standard “$\Lambda$CDM” model of cosmology describes a Universe that is spatially flat, dominated by cold Dark Matter (CDM) to enable formation of gravitationally bound structures like galaxies at early times, and a negative pressure term that drives the acceleration of expansion in later epochs, a ‘cosmological constant’ $\Lambda$.  When studying the late-time expansion of the Universe driven by $\Lambda$, attributed to an unknown force called `dark energy' \citep{Riess98, Perlmutter1998}, the parameter we are most interested in constraining is the expansion rate of the Universe in the present epoch: the Hubble constant $H_\mathrm{0}$. GW cosmology with standard sirens can be viewed as a population inference problem: the parameters that describe the cosmological model become population hyperparameters: specifically, information about the luminosity distance-redshift relation predicted by the $\Lambda$CDM cosmological model is embedded in the the redshift distribution of GW mergers that can be measured indirectly, while the luminosity distance distribution can be inferred directly from the GW signals themselves.

GWs have the potential to resolve one of our major outstanding mysteries about the $\Lambda$CDM cosmological model. Our current precision measurements of $H_0$ are in a statistical `tension' of around $4\sigma$: on one hand, measurements that extrapolate the present day expansion rate from fitting the temperature fluctuations of the CMB with a model that derives from the standard $\Lambda$CDM cosmological model derive a value of $H_0 = 67.8\pm0.9\,\mathrm{km\,s^{-1}\,Mpc^{-1}}$ \citep{Planck}. Supernovae, specifically the standardizable candle thermonuclear Type Ia supernovae (SNe Ia), can be used as part of the cosmic distance ladder to derive a much higher value, $H_0= 73.0\pm1.4\,\mathrm{km\,s^{-1}\,Mpc^{-1}}$ \citep{Shoes2022}. It remains an open question as to whether this tension arises from some unaccounted for systematic error associated with the cosmic distance ladder or intrinsic to SNe Ia themselves, or whether our standard cosmological model -- $\Lambda$CDM -- is incomplete. GWs offer an independent probe of $H_0$ that can potentially be used to resolve this tension, as they do not require any tethering to the cosmic distance ladder.

GW cosmology methods are split into categories that are delineated by the mechanism by which one determines the redshift of the gravitational wave source: 
\begin{itemize}
    \item \textit{Bright sirens} have identifiable electromagnetic counterparts that can be associated with individual host galaxies from which a confident and precise redshift estimate can be obtained. We have only one unambiguous coincidence between a GW event and an EM transient that allows for host galaxy identification, GW170817.
    \item The \textit{dark siren} technique determines a much broader redshift distribution by assigning association probabilities between a GW event and each catalogued galaxy within the localization region from which it has originated. There are numerous systematic effects that must be considered in the dark siren method, ranging from the methodology to rank galaxies that are most likely to have hosted a GW event \citep{Gair_2023}, to catalog incompleteness \citep[e.g.][]{Gray2022}, to peculiar velocity effects \citep{Nicolaou_pecvel}. 
    \item The \textit{spectral siren} technique\footnote{The `spectrum' in question when referring to the spectral siren technique is the `mass spectrum', i.e. the mass distribution}, which uses only gravitational wave data to obtain an estimate of $H_0$ by using information about the overall rest frame mass distribution of CO mergers. Here, the redshift information is implicitly contained in the offset between the rest and source frame masses, as the masses observed at the GW detector are effectively redshifted by a factor $M_s = (1+z)M_r$. There are several sub-categories of spectral siren techniques, however all are based on using gravitational wave data alone to determine $H_0$.
    \item The \textit{Love siren} technique and its variations use information about the equation of state of neutron stars, obtained through information encoded in the GW waveform through phasing terms introduced by tidal deformability to evaluate the redshift of the source, again through relating the rest and source frame masses using the redshift relation above. Love siren analyses cannot be performed at current detector sensitivites, as they require a large population of BNS mergers and confident measurements of tidal deformability.
\end{itemize}
A an overview of these techniques can be found in \cite{mastrogiovanni2024GRcosmoreview}. 

\begin{figure}
\centering
\includegraphics[width=\columnwidth]{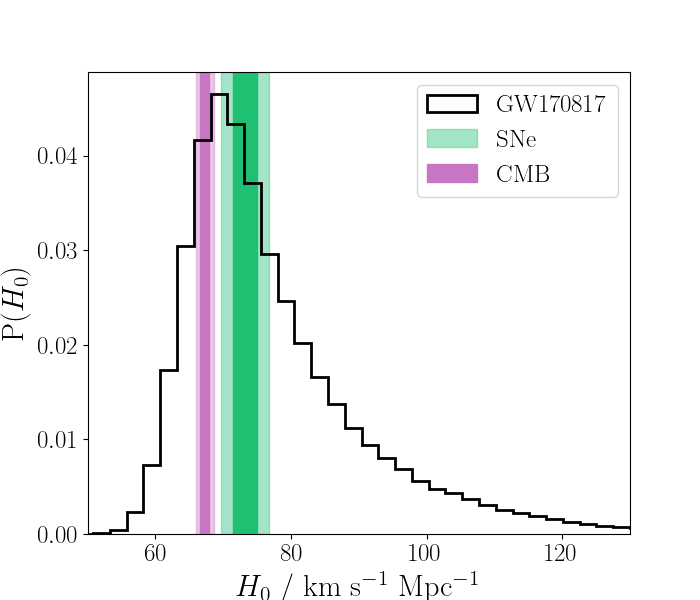}
\caption{Bright Siren measurement of $H_0$ from \citep{GW170817_cosmo} (black histogram) compared to SNe Ia constraint (green, \citep{Shoes2022}) and CMB constraint (purple, \cite{Planck2020}). Posterior PDF obtained from \url{https://dcc.ligo.org/LIGO-P1700296/public}}\label{fig:H0_bright}
\end{figure}

The present best constraints on $H_0$ from GW standard siren analyses use GW170817 as a bright siren, giving $H_0=70^{+12}_{-8}\,\mathrm{km\,s^{-1}\,Mpc^{-1}}$ (Fig \ref{fig:H0_bright}, \citep{GW170817_cosmo}). The uncertainty is dominated by there being only one observation from which a measurement can be made, and the degeneracy between the inclination angle of the binary and the luminosity distance distribution. This uncertainty can be effectively reducing by using late-time radio observations the radio jet produced by GW170817 to better constrain $d_L$ \citep{Hotokezaka2019} giving $H_0 = 86.3^{+4.4}_{-4.3}\,\mathrm{km\,s^{-1}\,Mpc^{-1}}$, as the main source of uncertainty in the measurement arises from the degeneracy between $d_L$ and $\iota$. Other analyses also fold in systematic effects like the peculiar velocity of NGC4993 \citep{Mukherjee2021}. In GWTC-4, $H_0$ is estimated by combining the posterior distributions on $H_0$ from the GW170817 bright siren with a dark siren analysis to give $H_0=76.6^{+13.0}_{-9.5}\,\mathrm{km\,s^{-1}\,Mpc^{-1}}$ (median and 68.3\% CI, \cite{GWTC4_cosmo}). A comprehensive overview of state-of-the-art standard siren measurements of $H_0$ using the techniques described above can be found in \citep{Jin2026}.



\section{Individual GW events beyond vanilla waveforms}\label{sec:nonvanilla}
As we saw in Section \ref{sec:waves}, GW events carry information about the physical properties of the merging COs. As our GW detectors increase in sensitivity, we are increasingly able to measure properties of CBCs beyond first-order effects. Studying these so-called `non-vanilla' events, where waveforms carry with them the imprints of spin precession, orbital eccentricity or higher order modes, we can learn more about the genealogy of CBCs across the Universe.

One of the big questions that has emerged as our gravitational wave catalogs have grown is \textit{`do we only detect circularised, non-precessing binary systems?'} Identifying and measuring the eccentricity of merging COs\footnote{BNS systems are generally expected to circularise and align their spins as forming a BNS system that mergers in a Hubble time through dynamical capture is vanishingly rare}, or identifying evidence of spin precession, would allow us to obtain a more rigorous insight into their evolutionary history. This is because these effects are expected to arise almost exclusively from systems that form dynamically \citep{Rodriguez2016_mergerrates, tagawa2023}. A measurement of either eccentricity or precessing spin complements constraints on the individual object spins, and the effective spin of the binary system, building a detailed picture of the likely formation scenario that gives rise to a GW event. 

Understanding these effects requires two things: Robust and efficient to compute waveform models that incorporate precessing spin, eccentricity or higher-order modes, and an abundance of GW data that can be confronted with these models. 

\subsection{Spin precession}\label{sec:spinprec}
The development of computationally-efficient waveform models that incorporate the effects of spin precession has allowed detailed studies of precessing spins in the handful of individual events where there is evidence for misaligned spins (the strongest candidates being GW190412 \cite{GW190412, Hoy2022}, GW191109\_010717 and GW200129\_065458 \cite{GWTC3}). However, contamination from transient glitches in detector data \citep{Payne2022,  Udall2025} and waveform systematics make conclusively detecting spin precession challenging. The most convincing evidence that we have detected a GW event with significantly precessing spins is GW200129. This event coincided with substantial excess noise in the Livingston interferometer. However, numerous works have continued to identify relatively strong evidence for precession \citep{Payne2022, Macas2024, Hannam2022, 4OGC}. However, there is some suggestion that waveform systematics must be carefully considered when measuring spin precession: for example, \cite{GWTC3} found evidence for precession when using the IMRPhenomXPHM approximant, but not with the SEOBNRv4 waveform model. Although tests for model misspecification have been suggested for GW astrophysics \citep{Romero-Shaw:2022ctb, Miller:2026buq}, results such as those above must be interpreted with appropriate caution.

\subsection{Eccentricity}
\begin{table*}[ht]
\centering
\caption{Binary black hole events with reported evidence for non-zero orbital eccentricity. Eccentricities are quoted at a gravitational-wave reference frequency of 10\,Hz ($e_{10}$) unless otherwise noted. Quoted uncertainties correspond to 90\% credible intervals.}
\label{tab:eccentric_events}
\begin{tabular}{lll}
\hline\hline
Event & Measured eccentricity & Reference \\
\hline
GW190521          & $e_{10} \geq 0.1$          & \cite{Romeroshaw2020} \\
GW190521          & $e = 0.69^{+0.17}_{-0.22}$                                  & \cite{Gayathri} \\
GW190620          & $e_{10} \geq 0.05$                                     & \cite{Romero-Shaw_2021} \\
GW191109          & $\geq 70\%$ posterior support at $e_{10} \geq 0.05$                         & \cite{Romero-Shaw_2022} \\
GW190701          & $e_{10} = 0.35^{+0.32}_{-0.11}$                                             & \cite{Gupte2025} \\
GW200129          & $e_{10} = 0.27^{+0.10}_{-0.12}$ to $0.17^{+0.14}_{-0.13}$           & \cite{Gupte2025} \\
GW200208\_222617  & $e_{10} = 0.35^{+0.18}_{-0.21}$                                             & \cite{Gupte2025} \\
GW190929          & Residual support for non-zero $e_{10}$                                      & \cite{Planas2025} \\
\hline\hline
\end{tabular}
\end{table*}

There is moderate evidence for the detection of eccentric mergers (Table \ref{tab:eccentric_events}). The degeneracy between the effects of spin precession and eccentricity on the observed waveform makes inferring the existence of both eccentricity and spin precession simultaneously not possible at current detector sensitivities \citep{Romeroshaw2023}. This is exacerbated in higher mass CBCs: although in principle the enhanced signal amplitude relative to the noise floor compared to signals from lower mass systems should allow for information about higher-order contributions to the waveform amplitude and phase evolution to be extracted, there are fewer cycles present in the LIGO band that can be used to search for the phasing and amplitude modulating effects of these phenomena. For example, in events like GW190521, novel techniques must be used to disentangle these effects \citep{Miller2024}. However, other works have claimed that it is not possible to distinguish whether this event is a circularised, spin-precessing system or an eccentric merger \cite{Romeroshaw2020} due to so few GW cycles being present in the LIGO band. 

\subsection{Higher order multipole emission}
Beyond these additional physical parameters that describe the compact binary system, another higher-order contribution to the frequency and amplitude evolution of the GW signal is the presence of higher-order modes. As noted earlier, the (2,2) quadrupolar mode dominates the detected GW signal in current GW detectors, however if CBCs are sufficiently asymmetric in mass ratio, higher order modes make substantial contributions to the observed waveform. The GW events GW170729 \citep{Katerina2019}, GW190412 \citep{GW190412} and GW190814 \citep{gw190814} - all of which show evidence for higher order modes. As GW detectors become more sensitive, the inclusion of higher order modes in analyses is now commonplace \citep{gwtc2} and neglecting to include multipole emission can have non-negligible impacts on analyses \citep{astrocalibration}. Mergers with very asymmetric masses are expected to have higher order multipoles that are more readily measured, which is why the most asymmetric CBC systems GW190412 ($q\sim0.28$) and GW190814 ($q\sim0.112$) exhibit the most conclusive evidence of higher order modes. However, the model misspecification problem also presents a particular challenge in disentangling the qualitatively similar effects of eccentricity and higher order modes on CBC waveforms as observed by LIGO. For example, there is disagreement in the literature as to whether GW190521 exhibits evidence of eccentricity or precession, as discussed above, or whether the signal can be better explained by one that has a measurable subdominant multipole component \cite{Hoy2022}.

\subsection{Methodological challenges}

Many searches for these phenomena focus on detailed study of events already identified by GW searches. These searches are designed to detect signals that arise from circularlised CO binaries.\footnote{GW search template banks usually assign spins to be aligned with the orbital angular momentum of the binary, with magnitudes drawn from a uniform prior.} However, detection relies on there being a substantial overlap (usually $\geq 90\%$ between the template and observed waveform). If the effects of these higher order effects are substantial enough to degrade this overlap, signals that arise from these systems will be missed in standard matched filter searches. Searches that can detect signals from eccentric or precessing systems can be performed using minimally modelled techniques that are more sensitive to signals exhibiting these effects \citep[e.g. ][]{Eccentric_search}, or using templates that are sensitive to eccentricity \citep{Phukon:2024amh,Wang_2025} or precession \citep{Precessingsearch}. These searches will become more important as increased detector sensitivies amplify the potential template mismatch with eccentric and precessing signals. Recent work has shown that incorporating the effects of spin precession in NSBH searches can more tightly constrain the merger rate in the local universe \cite{Harry2026}, showing that if there is no preferred direction for component spin, the overall NSBH merger rate is $\sim16\%$ smaller than that reported in \citep{GWTC3_pop}.

\section{Coincident multimessenger observations}\label{sec:multi}
The earliest GW detections prompted feverish searches for electromagnetic counterparts from astronomers not content to extract information from gravitational wave data alone. Such searches were well-motivated, with a well-established theoretical connection between BNS mergers and short GRBs \citep{Berger2014}.

The potential for serendipitous discovery of a BNS and its short Gamma-Ray Burst (sGRB) counterpart motivate low-latency searches. These pipelines operate alongside matched filtering searches and identify potential coincidences between GW triggers reported to GraceDB and high-energy transients reported by gamma-ray telescopes like Fermi-GBM, Swift and (in the past) INTEGRAL. The Rapid on-source VOEvent Coincident Monitor \citep[RAVEN][]{urban2016} is used to search for such coincidences in sky location and time. The Low-Latency Algorithm for Multi-messenger Astrophysics \citep[LLAMA][]{LLAMA} pipeline performs a similar search for coincidences between GW events identified in real-time by the LVK and high-energy neutrinos (HEN).

Beyond low-latency searches, there are multiple `targeted searches' that ingest catalogs of transients and re-analyze gravitational wave data to investigate whether any associated signal has been missed in the `all sky' searches that target any and all GW events. These searches are conducted using both modelled and unmodelled search methods using the PyGRB \citep{pygrb} and X-Pipeline \citep{Sutton_2010}. No significant evidence for GW signals associated with the GRBs has been found, nor for a population of unidentified subthreshold signals, since the beginning of the Advanced detector era either by LVK searches or external groups \citep{O3aGRB, O3bgrb}. These searches have focused predominantly on sGRBs; however, recent work \citep{Sinha2026} studying the potential connection between long GRBs (lGRBs) and GW-emitting CBC events \citep[e.g. motivated by recent identification of kn like emission associated with some lGRBs][]{KN_lgrb, lGRB_KNEmodelling} also yields a null result. As GW detectors probe a greater volume of the universe, it is likely that such deep targeted searches prove more fruitful, given the vast majority of GRBs we detect originate beyond the horizon of our current detectors. 

Nearby core collapse supernovae are also of great interest as a potential multimessenger GW source, although modelling of the GW emission is highly uncertain and the weak GW emission is only detectable for the most nearby supernovae \citep[e.g. ][]{Gossan2016, Marek2021}. Multimessenger SN observations can probe the processes that occur during core-collapse, with GWs providing smoking-gun evidence for phenomena such as bar-mode instabilities, standing accretion shocks, and QCD phase transitions in the protoneutron star \citep[See][ for a detailed review]{DAWES_burst}. Searches for GW emission from CCSNe out to $\sim 20\,\mathrm{Mpc}$ (which encompasses the Virgo group) have been routinely carried out since O1 \citep{Marek2024}. The recent detection of SN 2023ixf in M101 - one of the closest SN II ever observed - was used to constrain the GW luminosity of the event and the protoneutron star ellipticit \citep{SN2023ixfGW}.

Fast Radio Bursts (FRBs) may also have origins in CO mergers. The proposal that these highly energetic bursts of coherent radio emission lasting a few tens of $\mathrm{ms}$ could be emitted by merging neutron star or neutron star-black hole binaries had been explored thoroughly from a theoretical perspective \citep[e.g. see the review of][]{Zhang2023_rev}. Models predicting the emission of FRB-like emission from merging neutron stars at times ranging from several hours before to days after merger through mechanisms like magnetospheric interactions and magnetic braking, or the shedding of magnetic fields by hypermassive, highly magnetised `magnetar' remnants that undergo gravitational collapse as magnetic braking spins down the remnant and differential rotation quenches. In \cite{O3a_FRB} paper, the LVK perform a deep search for gravitational wave transients associated with FRBs detected by the CHIME telescope. No significant evidence of any GW events associated with these FRBs was found, although given the large uncertainties on the distance to many of the FRBs, it is not possible to entirely exclude the possibility of GW emission from NSBH mergers. This search strategy is targeted to GW emission that occurs close to the time at which the FRB is emitted, within a window of $\pm11\,\mathrm{seconds}$. An independent search in \cite{Wang2022} using the 4-OGC \citep{4OGC} GW transient catalog also returned a null result. 

\subsection{The first binary neutron star merger: GW170817}
On August 17 2017 at 12:41:04 UTC, the first binary neutron star merger was detected with a combined SNR of 32.4 and $\mathrm{FAR}=8\times10^4\,\mathrm{yr^{-1}}$. GW170817 originated from binary neutron stars with individual masses between $1.17 -1.60\,\mathrm{M_\odot}$, with the total mass of the system $M_\mathrm{Tot} = 2.74^{+0.04}_{-0.01}$ \citep{GW170817_props}. The signal was initially detected as a single-detector event present only in the LIGO-Hanford detector due to a significant glitch in the LIGO-Livingston detector just 1.1s before the merger, and the sky location of the source being in a null of the Virgo detector. The absence of any clear signal in the Virgo detector ($\mathrm{SNR}\sim 2$) was critical in reducing the sky localization uncertainty of the event to only $\sim 30\deg^{2}$ in low-latency \citep{GW170817}.

From the GW signal, it was possible to not only measure detailed information about the mass of the merging BNS system, but also to constrain features of the internal structure of the merging neutron stars \citep{GW170817_props}. The most well-constrained parameter is the chirp mass, $\mathcal{M} = 1.188^{+0.004}_{-0.002}M_\odot$, while the estimates of the individual masses are less precise owing to degeneracies in the mass ratio and the aligned spin components that depend strongly on the priors chosen for the spin components. A low-spin prior, motivated by stringent constraints on realistic NS EOS suggests lower individual NS masses and more equal mass ratios, with aligned spin components consistent with zero and $\chi_\mathrm{eff} \in [-0.01, 0.02]$ \cite{GW170817_props}. From the tidal deformability, a constraint on the neutron star radius of $r_\mathrm{NS}<13\,\mathrm{km}$ \citep{Raithel2018} could be obtained. This measurement was competitive with existing constraints on the NS EOS as measured from electromagnetic observations \citep{Raithel2019}, and would later yield joint constraints on the EOS and the expansion rate of the Universe ($H_0$) when combined with observations with the NICER x-ray instrument \cite{Dietrich2020}. 

$1.7\,\mathrm{s}$ after the merger that produced GW170817, the \textit{Fermi} Gamma-ray Burst Monitor (\textit{Fermi}-GBM) detected a gamma-ray burst -- GRB170817A. The detection was announced via GCN just 14 seconds later. The same GRB was detected by the \textit{INTEGRAL} satellite by the ACS of the SPI instrument via a search triggered by the \textit{Fermi} GCN and the LIGO-Virgo report of the GW event. The GRB event was characterised as a `short' GRB, with $T_\mathrm{90} = 2.0\pm0.5\,\mathrm{s}$\footnote{the time interval over which 90\% of the burst fluence is accumulated, in this case over energies $50-300\,\mathrm{keV}$}. The apparently coincident detection of a GW signal consistent with a BNS merger together with the sGRB, long theorised to originate in binary neutron star mergers, triggered expansive observing campaigns across the world, searching for other electromagnetic counterpart emission across the electromagnetic spectrum with ground- and space-based facilities. Detailed accounts of identification of the electromagnetic counterpart can be found in \cite{gw170817_mma} and the subsequent multi-wavelength follow-up effort in \cite{Margutti}. The event has become arguably one of the most well-observed astronomical phenomena in human history.

\subsection{Methods to associate GW and EM transients}
The association between GW170817 and GRB170817A provides a valuable exercise in quantifying the degree to which we believe that any two astrophysical transients are truly associated with one another. In the frequentist context, this requires us to establish whether there is sufficient evidence that we can reject the null hypothesis that the sGRB and GW detection events are independent of one another. In the LVK discovery paper, this is done by breaking this null hypothesis down into the temporal and spatial associations, and evaluating the probability of spatial and temporal coincidence separately before recombining them into an overall association probability.

The temporal association is more trivial: under the null hypothesis, the GW and sGRB detection events are independent Poisson processes, the probability of association is $P_\mathrm{temporal} = 5.0\times10^{-6}$, corresponding to $4.4\sigma$ significance in Gaussian statistics \citep{GW170817}. The temporal association alone, then, does not clear the threshold of $\sim 5\sigma$ for a truly unambiguous detection. However once folding in a quantification of the spatial agreement between the LIGO-Virgo sky localization of the GW source, and the Fermi GBM sky localizations, the significance improves substantially. In this specific case, this is done by defining a statistic that incorporates the posterior probabilities on the sky localization evaluated over a HEALPix pixelization of the sky that is compared to a background distribution that is obtained by bootstrapping a sample of 164 sGRBs that have been localized through targeted searches by GBM in an effort to quantify the morphology and size distribution of GBM sGRB localizations. The combined probability of chance association is then found to be $P_\mathrm{assoc}= P_\mathrm{temporal}\times P_\mathrm{Spatial} = (5\times10^{-6})\times 0.01 = 5\times10^{-6}$, rejecting the null hypothesis with a Gaussian equivalent significance of $5.5\sigma$ - an unambigouous discovery \citep{GW170817} by the standards of physics. The frequentist approach is one that is often taken in the literature when evaluating putative multimessenger detections, however care must be taken in this approach as it requires one to carefully quantify event rates (for example, the sGRB rate in the example above), instrument fields of view, and in the case that there are long delay times between a GW event and its claimed coincident multimessenger counterpart, to factor in the effects of Earth's rotation. 

This frequentist approach, particularly in the bootstrapping of the Fermi-GBM spatial response, could be seen as undesirable. For this reason, an alternative method to evaluate quantify the probability of association between GWs and multimessenger signals has emerged. The Bayesian Coincident Detection Criterion (`AC/DC') was introduced in \cite{Ashton_2018} and defines the `posterior overlap integral' that can be used to calculate the odds ratio between the hypothesis that a GW-EM coincidence arises from a common source as opposed to a random chance coincidence of two unassociated events. This formulation incorporates information about the signal hypothesis directly as opposed to merely rejecting the noise hypothesis based on Gaussian statistics. In the case of GW170817+GRB20170817A, the AC/DC method yields a total odds ratio of $\mathcal{O}_{C/SS}(D_\mathrm{GW}, D_\mathrm{EM}) \geq 10^{6}$, considered to be decisive evidence that the observations arise from the same astrophysical event. The overlap integral has been applied in a number of other theoretical works that demonstrate how it can be used to validate coincidences between EM and GW events, as well as in detailed analysis of claims of EM counterparts to other gravitational wave events, including GW190521 \citep{Ashton2021} and GW190425 \cite{Magana2024} as well as assessing coincidences between non-GW transients \citep{Sarin2024}.

\subsection{GW150914 and the Fermi-GBM claim}
The first detection of gravitational waves immediately provoked astronomers to search for electromagnetic counterparts. This was in spite of a dearth of theoretical explanations for how such emission could be produced, as stellar mass black hole binaries that co-evolve are not expected to have a plentiful supply of matter that can interact to produce detectable radiation. A small handful of mechanisms were, however, proposed following the detection of GW150914 \citep[e.g.][]{Kotera2016, Loeb2016, Murase2016, Perna2016, Stone2017, Zhang2016}. The trigger that prompted the sudden rash of theoretical explanations for a connection between BBH mergers and electromagnetic emission derived from the apparent coincident detection of a low-significance gamma-ray burst candidate by the Fermi-GBM telescope. The gamma-ray emission lasted only $\mathrm{1\,\mathrm{sec}}$ with a fluence of $2-3\times10^{-7}\,\mathrm{erg\,s^{-2}}$, which when combined with the luminosity distance of the source suggests an isotropic equivalent energy of $E_\gamma \sim 5\times10^{48}\,\mathrm{erg}$. The proposed mechanisms have varying explanations as to how a pair of merging black holes can launch a relativistic jet, ranging from the black holes involved in the merger being charged, to more conventional mechanisms that result in the extraction of rotational energy, or large magnetic fields or accretion rates from a disk that surrounds the merger remannt. The idea that the potential GRB detected by Fermi-GBM (which was not detected by any other gamma-ray observatories) and GW150914 are associated has largely fallen out of favour. Nevertheless, theoretical works that posit mechanisms by which BBH mergers may produce electromagnetic emission have become more widely accepted in more recent years, as evidence mounts for a population of BBH sources that have formed through dynamical interactions, potentially in gas-rich environments such as AGN. 

\subsection{GW190425 and the FRB20190425A claim}
The second BNS merger GW190425 \citep{GW190425} offered an exciting opportunity to search for electromagnetic counterparts. While enthusiasm for a second multimessenger event was buoyed by the detection of the multiwavelength emission from GW170817, the characteristics of GW190425 made for a much more challenging search. The event was identified initially as a single-detector event and could only be localised to within $\sim 1\times10^{4}\deg^2$ (around a third of the whole sky), and located around four times further in luminosity distance with $d_L\sim 160\,\mathrm{Mpc}$.

An popular method to search for associations between transients is to take an approach of catalog matching. \cite{Moroianu2023} use this method to identify a putative association between GW190425 and a CHIME FRB event FRB20190425A, which occurred 2.5 hours after the GW event was identified. Coincidences are identified by searching for CHIME FRBs that occur in a 26-hour time window about the time of merger. The window that encompasses two hours pre-merger to 24 hours post-merger, motivated by the wide variety of potential emission mechanisms that can result in coherent radio emission from compact binary events \citep{Chu2016, Zhang2023_rev}. The claim of an association between GW190425 and FRB20190425A relies on the rejection of the null hypothesis: that the two events are both astrophysical but originate from different, independent sources. The method is similar to that used to validate the association between GW170817 and GRB170817A \citep{gw170817_mma}.

There are two complicating factors that make interpreting the null hypothesis more challenging in the case of a GW-FRB association. One must account for the relationship between the redshift (and hence luminosity distance) and dispersion measure of the source, as the distance to an FRB cannot be measured directly. Moreover, in the case of \cite{Moroianu2023}, the long delay time between the GW event and the CHIME event ($\mathcal{O}(10^3\,\mathrm{s})$) means that the Earth's rotation must also be accounted for in computing the background rate of CHIME events. The coincidence is deemed marginally significant at $2.8\sigma$. However, explaining the coincidence requires extreme assumptions about the NS EOS that could result in the collapse of a hypermassive neutron star as large as the remanant of GW190425 ($M_\mathrm{Tot}\sim 3.4\,\mathrm{M_\odot}$) so long after merger \citep{Ai2024}. Even with mass loss of $\sim1\,\mathrm{M_\odot}$, the remnant would likely exceed the TOV maximum mass by a significant margin. 

Further investigation into the veracity of the claim was enabled by identification of the host galaxy of the FRB, UGC10667. This was done using both a probabilistic formalism \citep{Panther2023_frb} and later verified when analysis of the CHIME baseband data \citep{Bhardwaj2024} enabled localization of the FRB to within a few tens of arcminutes. Although observations of the host galaxy itself yielded little insight into the veracity of the association of the GW event \citep{Panther2023_frb}, a more refined analysis of the properties of GW190425 with restricted localization priors revealed inconsistencies with the orientation of the binary system in a more thorough analysis \citep{Magana2024}. Further optical followup published subsequently find no evidence for any kilonova emission associated with the BNS in the days that follow the merger \citep[][and see also systematic followup listed in Table \ref{tab:o3_surveys_uvoir}]{Smartt2024}. 

The significance of the association found by \citep{Moroianu2023} is strongly dominated by $P_\mathrm{DM}$, the chance probability that one finds a CHIME FRB with a DM that is more consistent with the redshift posterior of GW190425. FRB20190425 has an unusually low DM within the CHIME sample \citep[e.g. see Fig. blah of][]{Panther2023_frb}, placing it close to the LVK's O3 BNS detector horizon of $\sim 150\,\mathrm{Mpc}$ - for the same reason, the event was a stand-out in the deep search analysis \citep{O3a_FRB}. Based on the observational evidence above, and the extreme physics required for a hypermassive remanant to survive $2.5\,\mathrm{hr}$ before collapsing, it is thought that FRB20190425A and GW190425 are \textit{not} astrophysically associated. However, this coincidence claim reveals subtleties that must be considered carefully when making claims of astrophysical associations between GW events and putative counterparts.

\subsubsection{GW190521 and the AGN Flare claim}
GW190521 was detected by four real-time gravitational wave search pipelines on 21 May 2019 \citep{gw190521}. At the time, GW190521 was the most massive black hole binary system ever detected. The signal is consistent with the merger of a pair of black holes with component masses $m_1 = 85M_\odot,\, m_2 = 66M_\odot$. The simplest interpretation of the signal is that of a quasi-circular merger between two of the most massive black holes ever detected by the LVK. Black hole binaries with such large component masses cannot readily be explained through conventional stellar evolution models, so it has been widely accepted that the merging system responsible GW190425 formed through hierarchical mergers of dynamically formed black holes in a dense stellar environment. 

For this reason, the event has been reanalysed many times to search for evidence of any `smoking gun' in the gravitational wave signal that can reveal the evolutionary history of the binary black holes, for example, spin precession or any residual eccentricity that could provide more than circumstantial evidence for successive dynamical captures (see Section \ref{sec:nonvanilla}). 

 An interesting consequence of the formation channel proposed to explain GW150914 arises if the merger occurs in a gas-rich dense environment like the disc of an AGN. Various mechanisms for EM emission associated with such mergers have been proposed: The merging black holes or the merger remnant can interact with the gaseous accretion disk of the AGN through gas accretion or ram-pressure stripping of gas by the remnant as it is kicked through the disk after merger \citep{McKernan_2019}, or from the breakout emission from relativistic jets \citep{tagawa2023,Rodramirez2024}.

 Following GW190521, the Zwicky Transient Facility (ZTF) identified an AGN flare that was found to be consistent with emission expected from a BBH merger occuring in an AGN \citep{graham2020candidate, graham2023light}. The flare, named ZTF19abanrhr, stood out against the background of the AGN's intrinsic variability, and occurred around 34 days after GW190521 at a position that was evaluated to be spatially coincident with the GW event. The model invoked to explain the flare as an associated EM transient invokes the 'kicked Hill sphere' mechanism of \cite{McKernan_2019}, where gas that is gravitationally bound to the BBH merger remnant collides with gas in the AGN disk as the remnant recoils due to GW emission. The interaction between the kicked gas and the disk gas produces emission at UV and optical wavelengths, visible as a luminous transient that occurs many weeks after the merger, rising above typical AGN variability. 

 The AGN flare ZTF19abanrhr has never been conclusively associated with GW190521. Several works have disputed the claim on various grounds, including the large localization volume of the GW source \citep{palmese2021ligo} which leads to a conclusion that there is a 70\% chance of chance coincidence. Similarly, the result is disputed in \cite{Ashton2021} on the grounds that the common source odds range between 1-12 depending on waveform model selection, and in \cite{veronesi2025agn} based on a spatial correlation analysis. In spite of this, it has been used to produce posterior PDFs on the Hubble constant, highlighting an important opportunity \citep{Gayathri2020}. If we can reliably identify BBH mergers that are 'illuminated' by interactions with their merger environments, there is the potential to dramatically increase the number of `bright sirens' we can incorporate into our cosmology analysis (for example, GW190521 has been used as a bright siren to measure $H_0$, \cite{Mukherjee2020}). However, mis-identifications run the risk of introducing substantial biases and systematics into such measurements that must be accounted for: there is far more work to be done on methods to establish robust associations in scenarios like that proposed to explain GW190521+ZTF19abanrhr.

\subsection{Multimessenger sources in O3 and O4}
During O3, the dissemination of GW detections in real time through LVK Open Public Alerts (OPA) allowed teams of electromagnetic astronomers to perform systematic searches for EM counterparts to detected GW events. In this work, we have tabulated these surveys in Tables \ref{tab:o3_surveys_uvoir} and \ref{tab:o3_surveys_highE} based on \citep{Poggiani2024} and a thorough search of literature that post-dates it.
\begin{table*}
\centering
\caption{Summary of all-O3 systematic UVOIR follow-up programs for gravitational-wave electromagnetic counterparts.}
\label{tab:o3_surveys_uvoir}
\renewcommand{\arraystretch}{1.3}
\footnotesize
\begin{tabular}{@{}p{3.7cm}p{2.0cm}p{6.4cm}p{3.0cm}@{}}
\toprule
\textbf{Collaboration / Facility} & \textbf{Wavelength} & \textbf{Scope} & \textbf{Reference} \\
\midrule
\multicolumn{4}{@{}l}{\textit{Optical / Near-Infrared survey programs}} \\
\midrule
GRANDMA (O3a) & optical / NIR & First half of O3; 25-telescope global network & \citep{Antier2020a} \\
GRANDMA (O3) & optical / NIR & 49/56 GW alerts; $\sim$9000 deg$^{2}$ total coverage & \citep{Antier2020} \\
ENGRAVE & optical / NIR & VLT-led EU multi-facility collaboration; GW190814 flagship campaign & \citep{Ackley2020} \\
GROWTH / ZTF (S190425z) & optical ($g$, $r$) & First ZTF GW O3 alert; rapid response demonstration & \citep{Coughlin2019} \\
GOTO & optical ($L$-band) & 29 O3a GW triggers; $\sim$732 deg$^{2}$ avg.\ per event & \citep{Gompertz2020} \\
DDOTI & optical ($w$-band) & 26/48 unretracted O3 GW alerts; rapid wide-field response & \citep{Becerra2021} \\
SAGUARO (O3 follow-up) & optical & 17 GW events with CSS data; Mt.\ Lemmon 1.5\,m & \citep{Paterson2021} \\
J-GEM (galaxy-targeted) & optical / NIR & 23 GW events in O3 (11 BBH, 5 BNS, 4 NSBH, 1 MassGap, 1 Burst, 1 Terrestrial) & \citep{Sasada2021} \\
SkyMapper & optical ($i$) & Robotic 1.3\,m follow-up; AlertSDP pipeline, $i_{\rm AB} \approx 20$ mag & \citep{Chang2021} \\
ASAS-SN & optical ($V$, $g$) & 9 O3 events with $>$60\% NS probability &\citep{deJaeger2022} \\
ZTF / GROWTH (BNS+NSBH) & optical ($g$, $r$) & 13 BNS/NSBH triggers in O3; 340 photometric points, 64 OIR spectra, 3 radio epochs & \citep{Kasliwal2020} \\
GECKO & optical / NIR & KMTNet wide-field follow-up; first three BBH events in O3a (GW190408, GW190412, GW190503) &\citep{Kim2021} \\
ZTF + GOTO (O3b ejecta) & optical & Ejecta-mass constraints on O3b BNS/NSBH triggers (S191205ah, S191213g, S200115j, S200213t) &\citep{Coughlin2020} \\
Multi-survey systematic & optical / NIR & 653 candidates from 15 O3 mergers $\rightarrow$ 66 plausible KN after vetting & \citep{Rastinejad2022} \\
KN detectability post-mortem & optical / NIR & Detectability/recoverability assessment of O3 KN searches & \citep{Sagues2021}\\
ZTF (blind 23-month KN) & optical ($g$, $r$) & 23-month blind ZTF search for rapidly-evolving transients; KN luminosity function and rate & \citep{Andreoni2020} \\
ZTF / GROWTH (BBH AGN flares) & optical ($g$, $r$) & O3 BBH localizations cross-matched to ZTF AGN flares; 9 candidates (7 O3a + 2 O3b) & \citep{Graham2023}\\
Pan-STARRS (blind KN) & optical ($griz$, $w$) & Blind KN search 2019-Oct to 2022-Dec; 29\,740 transients reported; 175 within 200 Mpc; spans O3b through early O4a & \citep{Fulton2025} \\
\bottomrule
\end{tabular}
\end{table*}

\begin{table*}
\centering
\caption{Summary of all-O3 systematic high-energy (radio, X-ray, UV, gamma-ray, VHE) follow-up programs for gravitational-wave electromagnetic counterparts.}
\label{tab:o3_surveys_highE}
\renewcommand{\arraystretch}{1.3}
\footnotesize
\begin{tabular}{@{}p{3.7cm}p{2.0cm}p{6.4cm}p{3.0cm}@{}}
\toprule
\textbf{Collaboration / Facility} & \textbf{Wavelength} & \textbf{Scope} & \textbf{Reference} \\
\midrule
\multicolumn{4}{@{}l}{\textit{Radio}} \\
\midrule
Apertif (WSRT, GW190425) & radio (1.4 GHz) & 3-epoch radio follow-up of GW190425 BNS; demonstration of Apertif PAF capability & \citep{Boersma2021} \\
JAGWAR / VLA (GW191216\_213338) & radio (4--8 GHz, C-band) & Targeted VLA search for BBH-AGN candidate counterpart to GW191216\_213338 (originally identified as superevent candidate S191216ap) & \citep{2021ApJ...911...77B} \\
\midrule
\multicolumn{4}{@{}l}{\textit{X-ray / Ultraviolet}} \\
\midrule
Swift / XRT & X-ray (0.3--10 keV) & 18 GW triggers; $\sim$6500 pointings & \citep{Page2020}\\
Swift / UVOT & near-UV & 18 GW alerts; 424 deg$^{2}$; 27 sources + 13 cross-facility candidates & \citep{Oates2021} \\
Swift (GW + $\nu$ coincident) & X-ray + UV & 4 GW + IceCube coincident searches in O3 & \citep{Keivani} \\
Swift / BAT (sub-threshold targeted) & hard X-ray / soft $\gamma$-ray & 636 sub-threshold/significant GW candidates analyzed in BAT; 86 high-purity & \cite{Raman2025} \\
\midrule
\multicolumn{4}{@{}l}{\textit{Gamma-ray / Very-High-Energy}} \\
\midrule
Insight-HXMT/HE & 0.2--3 MeV & Targeted coherent search; O2 + O3a GW events & \citep{Cai2021} \\
H.E.S.S. (GW follow-up program) & VHE $\gamma$-ray ($>$100 GeV) & H.E.S.S. GW rapid follow-up strategies in O2 and O3 & \citep{Ashkar2021}\\
H.E.S.S. (BBH follow-up) & VHE $\gamma$-ray ($>$100 GeV) & 4 BBH mergers (GW170814, GW190512, GW190728, S200224ca); 35--75\% localization coverage & \citep{Abdalla2021} \\
CALET (CGBM + CAL) & 7 keV -- 10 TeV & X-ray/$\gamma$-ray counterparts during O3 & \citep{Adriani2022} \\
Fermi-GBM (precursor search) & gamma-ray (8 keV -- 40 MeV) & Targeted search for modulated $\gamma$-ray precursors to O2 + O3 CBC mergers & \citep{Stachie2022} \\

\bottomrule
\end{tabular}
\end{table*}

A number of searches identified transients of potential interest either through comprehensive follow-up strategies or dedicated targeting of specific events considered of interest (see Table \ref{tab:o3_coincidences} for papers associated with events reported in GWTC-2.1/3. Events not included in the catalog have been excluded from the table). The majority of these coincidences have subsequently been ruled out. GW190814 received possibly more than its fair share of attention due to its exceptional sky localization area ($\sim 18\deg^2$, \cite{gw190814}) in the context of O3 detections, and the mass posterior making the possibility that the event was a NSBH merger - and therefore many have produced detectable EM emission - a real possibility. In spite of this, no convincing candidate has been detected to date associated with this event. 

\begin{table*}
\centering
\label{tab:o3_coincidences}
\renewcommand{\arraystretch}{1.3}
\small
\begin{tabular}{@{}p{2.8cm}p{6.5cm}p{6.0cm}@{}}
\toprule
\textbf{GW Event} & \textbf{References} & \textbf{Comments} \\
\midrule
\multicolumn{3}{@{}l}{\textit{Confirmed events in GWTC-2.1 / GWTC-3}} \\
\midrule
GW190814 (UVOIR) & \cite{Andreoni2020, Ackley2020, Vieira_2020, Kilpatrick2021, Tucker2022, Morgan_2020, DeWet2021}& NSBH / BBH? Many UVOIR candidates including DG19wxnjc / AT2019npv; all ruled out, AT2019npv = SN Ib-like \\
GW190814 (Radio) & \citep{Alexander2021, Dobie2022} & ASKAP J005022$-$230349 and 75 VLA galaxy fields; no viable radio counterpart \\
GW190814 (X-ray) & \citep{Page2020} & Swift-XRT pointed observations; no viable X-ray counterpart \\
GW200105 & \citep{Anand2021} & NSBH; ZTF candidates all ruled out \\
GW200115 & \citep{Anand2021} & NSBH; ZTF candidates all ruled out \\
7 $\times$ O3 BBH\textsuperscript{a} & \citep{Graham2023, He_2025} & 7 ZTF AGN flares cross-matched with O3 BBH localizations; 4/7 rejected on re-analysis with extended ZTF data \citep{He_2025}; 2 of remaining 3 retain strong joint correlation \\
\bottomrule

\end{tabular}
\caption{Follow-up of potential EM coincidences with events confirmed in GWTC-2 and GWTC-3. GW190425 and GW190521 are discussed in detail in the text and are excluded.}
\end{table*}

The analysis of multimessenger searches performed during O4 are still largely ongoing. During O4, the sources that gained the most attention from the multimessenger community were largely those deemed to have involved at least one neutron star, however no convincing EM counterpart candidate has emerged from these searches. A summary of O4 events that provoked significant multi-messenger follow-up responses can be found in Table \ref{tab:o4_coincidences}. This table is considered to be relatively complete at the time of writing with regards to events reported in GWTC-4 \citep{GWTC4}, however this space is expected to rapidly evolve with the release of GWTC-5 \cite{gwtc5} and GWTC-6 in the coming year.
\begin{table*}
\centering
\caption{Follow-up of potential EM coincidences with O4 gravitational-wave events, for GW events confirmed in GWTC-4.0.}
\label{tab:o4_coincidences}
\renewcommand{\arraystretch}{1.3}
\footnotesize
\begin{tabular}{@{}p{2.0cm}p{2.0cm}p{1.5cm}p{2.7cm}@{}}
\toprule
\textbf{GW Event} & \textbf{GW Type} & \textbf{Coinc.\ Type} & \textbf{Key Reference(s)} \\
\midrule
\multicolumn{4}{@{}l}{\textit{Confirmed events in GWTC-4.0}} \\
\midrule
GW230518\_125908\textsuperscript{a} (S230518h) & NSBH (ER15, pre-O4a) & UVOIR & \citep{Paek2025, Ahumada, Pillas2025} \\
GW230529\_181500 (S230529ay) & NSBH (lower mass gap) & UVOIR; $\gamma$-ray/X-ray & \citep{Kunnumkai2025, Ronchini, Ahumada, Pillas2025} \\
GW230627\_015337 (S230627c) & NSBH (subthresh.) & UVOIR & \citep{Ahumada, Pillas2025} \\
GW230922\_020344 (S230922g) & BBH (high-significance) & Optical (AGN disk) & \citep{Cabrera} \\
\bottomrule
\end{tabular}
\end{table*}

\section{Fundamental Physics from CBC events}\label{sec:fundamental}
GWs enable us to test the fundamental properties of gravitation, providing access to the extreme gravity regime of the General Theory of Relativity (GR). This section does not contain an exhaustive compilation of results on tests of GR, but provides examples of the methods that can be employed to study fundamental physics using GWs and potential extensions of both the Standard Model of particle physics, and the $\Lambda$CDM cosmological model.

\subsection{Strong field tests of GR}
Despite the remarkable success of GR in predicting phenomena associated with CBCs, it may not be a complete theory of gravitation. This sentiment does not come from any experimental bases. Rather, GR yields a number of results such as the formation of singularities \citep{penrose1965singularity} that cannot be readily explained without the unification of GR with another pillar on which modern physics stands: quantum mechanics. Straightforward attempts to reconcile GR and quantum mechanics famously fail, motivating various attempts to quantize gravity using e.g. strings \citep{carlip2015QuantumGravity}. To test these theories we can search for deviations in GR at energies far below those at which the four fundamental forces are unified, thus giving insight into the fundamental theory of gravity \citep{shankaranarayanan2022modified}. GWs from CBC mergers can be usedn to test GR, since this is arguably the first astronomical observation of a system where the non-linear effects of gravity become important. This is due to the strong curvature and highly dynamical nature of the spacetime around COs during the late inspiral and the ringdown of the merger remnant. In addition, even if GR is correct at all astrophysically relevant scales probed by CBCs, the exercise of testing GR is still of critical importance: it provides a consistency check of our understanding of the physics involved in CBCs and GW emission.



The majority of GR tests performed with LVK data are null tests built around a common conceptual framework: the best-fit GR waveform obtained via parameter estimation is subtracted from the detector data, and the residual is tested for consistency with Gaussian noise \cite{passenger2025gaussianGR}. Departures from GR would appear as coherent structure in the residuals. This framework is applied across each phase of the signal. During the inspiral, the phase evolution of the waveform is compared against post-Newtonian predictions \cite{GWTC4parameterizedGR}. After merger, the ringdown of the remnant is tested for consistency with the ringdown expected from an astrophysical (Kerr) black hole \cite{gwtc4RemnantsGR}. The mass and spin that can be inferred from the remnant's ringdown is compared with the predictions of GR from the inspiral phase of the \cite{GWTC4GRtests1} of the signal. Inconsistencies between the inferences made in the different phases can point to one of two things: a departure from GR predictions, or some unknown artefact introduced by systematic uncertainties in calibration or an unaccounted for noise transient. 

In GWTC-3, tests of GR \citep{abbot2025gwtc3GRtest} consistently find no statistically significant deviations from GR across a range of inspiral, merger, and ringdown tests. Beyond tests of the orbital dynamics, GR also makes precise predictions about the nature of gravitational waves themselves. For example, GR predicts exactly two tensor polarization modes, whereas more general theories of gravity often predict additional scalar or vector polarization states \citep{eardley1973polarization}. A detection of such modes would be a smoking-gun signature of physics beyond GR. No such detection has been made \citep{cheung2026GRpolarizations}.

We have already alluded to a critical challenge in all tests of general relativity: distinguishing beyond-GR signatures from noise artifacts, waveform systematics, or unaccounted for astrophysics \cite{gupta2024violations}. These possible false violations of GR are already relevant, e.g. in \cite{maggio2023grtest}, where they find the event GW200129\_65458 shows strong violations of GR. This is interpreted as either a data-quality issue or waveform systematic from mis-modelling of the spin precession. This underscores the importance in producing robust template BBH waveforms including spin precession, higher order multipoles, and eccentricity \cite{pang2018higherordermodesGRviolation}, and developing stringent methods to explore model misspecification in the Bayesian inference framework.

On the other hand, some argue that to find true deviations from GR we need to go beyond null tests, and look for specific deviations in GR predicted by alternate models for gravity. The production of NR waveforms in theories of modified gravity - which is essential to perform fully consistent tests of non-GR theories rather than merely parameterised ones - remains in its infancy. Certain classes of modified theory, notably those involving spontaneous scalarization of compact objects \cite{doneva2024scalarization}, predict genuinely non-perturbative effects during merger that cannot be captured by a perturbative expansion around a GR background, making dedicated NR simulations in the modified theory an unavoidable requirement \cite{okounkova202NR}.

\subsection{Speed of gravity}
A fundamental prediction of the General Theory of relativity is that the speed at which gravitational waves propagate is equal to the speed of light in a vacuum. This is a non-trivial fact that is violated by many different models of gravity beyond GR. For example, theories of massive gravity have been proposed to explain physics such as the accelerating expansion of our Universe \cite{deRham2014gravity}). The strongest constraint on $v_g$ has been derived from the multimessenger detection of GW170817, constraining $v_g=c$ within a factor of $\sim10^{-15}$ \citep{GW170817_cosmo}. Results from GWTC-4 investigate modified GW propagation finding results that are consistent with GR predictions that $v_g = c$ \citep{GWTC4_cosmo}.
\subsection{Dark matter and exotic compact objects}
As we have previously discussed, the standard model of cosmology requires the matter content of the universe to be dominated by `cold dark matter', or CDM. While it is clear that dark matter (DM) is phenomenologically required to explain the emergence of large scale structure in the Universe, as well as the rotation curves of galaxies and observations of displaced x-ray gas in galaxy haloes, the nature of DM is uncertain. 

One proposal is that DM is comprised of exotic, ultra-light bosonic particles. GWs offer a way to search for such exotic matter through the phenomenon of superradiance, where the gravitational coupling of the bosonic wave to a BBH merger remnant. The bosonic field can then extract energy from the rotational energy of the BH, leading to a lower mass BH with reduced spin. The majority of searches for vector and scalar boson clouds around BHs have returned null results \citep[e.g.][]{}, except for very recent very tentative evidence reported by \cite{roy2026scalar} that there could be a light scalar field of mass $m\sim 10^{-12}\rm{eV}$ ($\ln(\mathcal{B_{\rm vac}^{\rm env}})\approx 3.5)$ in the environment of GW190728. 

An alternative explanation for DM is the presence of a hitherto-undetected population of sub-solar mass (SSM) compact objects that formed in the early Universe from the collapse of primordial overdensities. These so-called `primordial black holes' differ from the known population of sub-solar mass compact objects (white dwarf stars) as they are substantially more compact, and so their mergers are detectable by ground-based laser interferometers. There is also a substantial interest in SSM mergers from an astrophysical perspective, as there have been several mechanisms posited for how such objects can form through processes associated with conventional stellar evolution, and how such mergers may produce observable EM signatures \citep{MetzgerSSM, ChenSSM}. Searches for SSM GWs using template banks \citep{Chad_SSMbanks} designed to matched-filter data to identify candidate signals have so far returned null results, with no SSM events identified. However, at the time of writing one compelling candidate SSM event \citep{SSM_o4c} was reported during the final weeks of O4. The most recent SSM searches by the LVK (based on data from O3) constrain the the fraction of dark matter that is comprised of primordial black holes, finding $f_\mathrm{PBH}\geq0.6$ can be excluded at 90\% confidence if the PBH masses are monochromatically distributed, however a broader PBH mass distribution cannot rule out $f_\mathrm{PBH}=1$ \citep{O3aSSM, O3bSSM}. Similar constraints are found by groups performing SSM searches on LVK data released to the public \citep[e.g.][and references therein.]{Kacanja_2025}



%

\section{Conclusions and Future prospects for CBC populations as a messenger}\label{sec:future}
The rapid development of GW astrophysics as an observational science has given us a taste of the discoveries that can be made using this unique messenger. In the somewhat biased spirit of this review, we briefly preview two potential avenues for GW discovery that exploit the hierarchical nature of the GW detection and parameter estimation problem. These future prospects should emphasise the power of GW discovery in not just revealing some of the most extreme individual objects in the Universe, but how a combined population can be more than the sum of its parts as we probe the edges of our understanding of the cosmos. 

\subsection{Post-merger GWs from neutron stars}

When two neutron stars merge, the system may promptly collapse to a black hole or, if conditions are right, form a differentially rotating remnant that survives the merger \citep{Sarin2021_rev}. Such a hyper- or supramassive neutron star can be supported well above the maximum mass of a cold, non-rotating star — by 30–70 per cent — through differential rotation and thermal support \citep{baumgarte00, Shapiro2000, Bauswein2013}. Over the first $\sim100\,\mathrm{ms}$, the post-merger remnant cools through processes including the emission of high-frequency ($\sim\,\mathrm{kHz}$) gravitational waves \citep{Hotokezaka2013}. Because the matter at the core of the newly formed post-merger neutron star is hot and ultradense, these signals probe a regime of the nuclear equation of state inaccessible to the inspiral, which constrains only the cold equation of state \citep{GW170817, GW190425}. An observation of post-merger GWs then offers the prospect of uncovering temperature-dependent phase transitions in some of the densest matter in the Universe \citep{Bauswein2019_phasetransition}. However, searches for post-merger emission from BNS mergers GW170817 \cite{GW170817postmerger, GW170817postlong} and GW190425 \citep{Grace2024} find no evidence for any signal, although \citep{VanPutten2018, Abchouyeh2023} claim that there may be features in the data consistent with post-merger GW emission. 

Our ability to extract this physics rests on waveform models conditioned on numerical-relativity (NR) simulations, which show that the dominant quadrupolar oscillation frequency correlates with the compactness and tidal deformability of the remnant \citep{Bauswein2012_eosdependence, Bauswein2012_NSproperties, hotokezaka13, Takami2014}. These simulations are finite and unevenly sample the parameter space of BNS masses, mass ratios and spins — there are, for example, too few incorporating rotational effects to fully characterise the relationship between the rotating maximum mass and the TOV mass \citep{Iosif2022, Cassing2024}. Analytic approximants calibrated to NR simulations achieve only modest overlaps \citep{Clark2016, Easter2020, Chatziioannou2017, Breschi2024}. Nonetheless, NR and analytic approximations of waveforms represent two complementary inference strategies: direct parameter estimation on a single loud event, or the use of quasi-universal relations between spectral features and physical parameters to constrain the equation of state more agnostically. There is opportunity here to investigate the application of machine learning -- in particular with techniques like symbolic regression -- to develop enhanced waveform models that can bridge the gap between computationally expensive and numerically unreliable NR simulations, and well-understood but oversimplified phenomenological models.  

The greater promise lies in populations. Individual post-merger signals will overwhelmingly fall below the threshold for confident detection, given the poor sensitivity of observatories in the kHz band, and may remain so even once Cosmic Explorer\citep{CEReitze}, the Einstein Telescope \citep{ET}, and proposed HF-optimized detectors like NEMO \citep{ackley20} or KAGRA-HF \citep{KAGRAHF} come online \citep{panther23}. Rather than waiting for a single loud event, information from many sub-threshold remnants can be combined coherently \citep{Yang2018, Criswell2022, Mitra2025}. By statistically inferring the fraction of mergers that avoid prompt collapse and folding in the neutron star mass distribution measured from the inspirals, an ensemble of only 25–35 events can constrain the maximum mass of hot, rapidly rotating neutron stars — and hence the TOV mass — to $\sim$10–20\% \citep{Panther26} and the NS radius to $\sim10\%$ \citep{criswell23}. These population-based approaches are comparatively forgiving of waveform model misspecification, since at low SNR a search is sensitive only to the loudest spectral feature, and points toward the statistical combination of weak signals as the most credible near-term route to the hot nuclear equation of state.

This population-driven approach can be complemented with an increased number of BNS inspiral observations to probe the boundary between the maximum neutron star mass and the minimum black hole mass. Electromagnetic observations of black holes in X-ray binaries have long suggested a dearth of compact objects between roughly $2.5-5M_\odot$, a gap whose lower edge is set by the maximum neutron star mass and whose existence bears on the supernova physics governing compact-object formation. Gravitational-wave populations now probe this transition from a complementary direction: analyses of the compact-binary mass distribution find evidence for a change in population near $2.4M_\odot$ \citep{GWTC4}, broadly consistent with predictions for the maximum neutron star mass, while events such as GW190814 and, more recently GW230529, challenge the notion that the region is empty. 

Combining the insights we can gain from the structure of the NS and BH populations in the lower mass gap and those from sub-threshold population methods like that proposed in \citep{Mitra2025,criswell23, Panther26} can yield valuable information about the maximum neutron star mass and minimum NS radius. As the catalogue of binary neutron star mergers grows, the combination of these population-level probes offers the most promising route to resolving whether the lower mass gap is a genuine feature of compact-object formation or merely an artefact of sparse sampling.

\subsection{Relativistic dynamics at the center of galaxies}

The dense stellar environments at the centres of galaxies — nuclear star clusters (NSCs) and the accretion disks of active galactic nuclei (AGN) — represent compelling but poorly understood sites for compact object mergers, and are among the most promising targets for future multi-messenger astrophysics. The fundamental challenge is that both the detailed dynamics of compact objects in these environments and the conditions under which their mergers produce observable electromagnetic transients remain deeply uncertain. Despite this, indirect evidence for an AGN contribution to the observed BBH population is accumulating, both through spatial localisation \cite{zhu2025evidence} and population-level BBH property analyses \citep{li2025agngwtc4}. Whether these signals are truly evidence of the AGN channel, or reflect systematic uncertainties in sky localisation and population modelling, remains an open question, and the expected electromagnetic counterparts - even if they exist — may be undetectable in the most massive, luminous AGN that dominate current catalogs \citep{veronesi2025constraining,grishin2024effect,Moncrieff2025}. Understanding when and whether BBH mergers in AGN produce observable flares, and how they connect to the underlying disk and NSC dynamics, is a key open problem for multi-messenger astronomy in the LIGO era and beyond.

The same galactic nuclear environments also host extreme mass ratio inspirals (EMRIs). An important and underexplored regime connects EMRIs to the broader stellar-mass BBH population: in AGN disks, stellar-mass black holes co-migrating toward the SMBH can be captured into a stellar mass binary, whose center of mass orbits the central SMBH. More broadly, BBH mergers occurring deep in the gravitational potential of the SMBH - whether driven by tidal capture \cite{chen2018bemri}, or disk migration \cite{peng2024fate} - represent the most complex realisation of a BBH coalescence conceivable: they involve large spins and hence significant precession, residual eccentricity, strong-field relativistic effects from the SMBH, and gas-driven orbital dephasing \cite{zwick2025environmental}, all simultaneously. Even if such mergers are rare in nature, they define an upper bound on the complexity of BBH coalescence's and serve as a valuable theoretical testbed for waveform modelling, and therefore crucial to understand before any claim of a deviation from GR to be accepted.

Beyond the gravitational dynamics, the AGN channel is a potentially rich multi-messenger source, arising from interactions of compact objects with the disc \citep{perna2021agn, gri21}, interactions of stars with the disc \citep{cantiello2021stellaragn, mckernan2022starfall}, and the rich variety of explosive transients these environments produce. Indeed, we may already be observing EMRIs electromagnetically, with disc-driven EMRIs plausibly connected to the growing population of observed TDEs and QPEs \citep{jiang2025agnqpetde, lyu2026wetemriscience}. Combining GW detections of these sources with their electromagnetic counterparts would greatly enhance our understanding of the high-energy physics at galactic centers, while simultaneously enabling precision tests of GR, and independent measurements of cosmological parameters via multi-messenger standard sirens \cite{laghi2021cosmology}.

\subsection{Probing the standard model of cosmology with GWs}
The rapid evolution of GW cosmology, and the development of methods to exploit the standard siren property of GW emission from CBCs has demonstrated that GWs are one of the strongest competitors for a method to break the Hubble tension \citep{GW170817_cosmo}. It is expected that with additional `bright siren' sources like GW170817, and improved sky localizations increasing the accuracy and precision of the dark siren method from the GW side, could yield inferences of $H_0$ that are accurate to around 10\% in the next decade \citep{hughes05}.

GW cosmology is presented with a challenge: to ensure that we understand and quantify our systematic uncertainties in our measurement of $H_0$, in much the way that has been done with other local-Universe probes of late-time expansion \cite{Scolnic2025}. This will require several things to occur: for us to develop robust methods to quantify biases that can be introduced into bright siren cosmology in the process of associating EM transients with GW events. This is especially important in an era that will be defined by the detection of up to 10,000 transients per night by the Vera Rubin Telescope \citep{Ivezic2019}. However, the potential for surveys like the Large Synoptic Survey of Space and Time (LSST) with the Vera Rubin Telescope will provide highly complete galaxy calatogs, dramatically improving dark siren measurements of $H_0$ \citep[e.g. ][]{Gray2023}. We are now ideally placed to use GWs to perform a rigorous test of our best theory that describes how our Universe formed the structures we admire with our telescopes today, and its ultimate fate into a distant and uncertain future.

\section{Acknowledgements}
FHP thanks Gavin Rowell for the invitation to write this review article. FHP is supported by a Forrest Research Foundation fellowship. JWNM is supported by funding from the Australian Government Research Training Program. The authors thank David Coward, Evgeni Grishin, Alessandro Trani, Paul Lasky, Eric Thrane, Ilya Mandel, Lachlan Passenger, Shun Cheung, Mallika Sinha, Alistair McLeod, Damon Beveridge, Chayan Chatterjee, Manoj Kovalam, Linqing Wen, Lilli Sun, Jade Powell, Simon Stevenson, Anais M\"{o}ller, Liana Rauf, Carl Blair, Chiara di Fronzo, Ju Li, Chunnong Zhao, Eric Howell and Bruce Gendre for useful discussions that have contributed to the perspectives presented in this work. FHP acknowledges colleagues from the LVK and beyond for the informative and interesting discussions we have had together over the past six years, in particular those who offered feedback to improve the manuscript: Chayan Chatterjee, Md Shaikh, Suvodip Mukherjee, Christopher Berry, Matthew Mould, Maciej Bilicki and Sharan Banagiri. The authors acknowledge the support of the ARC Center of Excellence for Gravitational Wave Astronomy (OzGrav), which has played an important role in supporting their work and enabling national and international collaboration in GW astrophysics. In particular, they acknowledge the director of OzGrav, Matthew Bailes, for creating a center that has uplifted the careers of numerous young scientists working on gravitational wave astronomy in Australia since 2017. This material is based upon work supported by NSF’s LIGO Laboratory which is a major facility fully funded by the National Science Foundation.
\bibliographystyle{elsarticle-num} 
\bibliography{bib_master_new}






\end{document}